\documentclass[fleqn,10pt,showpacs,final,twocolumn,nofootinbib,floatfix,superscriptaddress]{revtex4}

\usepackage{amsmath}
\usepackage{amssymb}
\usepackage{graphicx,psfrag,epsfig}
\usepackage{afterpage}
\usepackage{mathrsfs,simplewick}
\usepackage{color}

\newcommand{\be}{\begin{equation}}
\newcommand{\ee}{\end{equation}}
\newcommand{\bea}{\begin{eqnarray}}
\newcommand{\eea}{\end{eqnarray}}
\newcommand{\bean}{\begin{eqnarray*}} 
\newcommand{\eean}{\end{eqnarray*}}

\newcommand{\bm}[1]{\mbox{\boldmath $#1$}}

\def\cd{{\cdot}}

\def\!{\hat}
\def\slash{\rlap{/}}

\newcommand{\st}{{\scriptscriptstyle T}}

\def\slash#1{\setbox0=\hbox{$#1$}               
        \dimen0=\wd0                            
        \setbox1=\hbox{/} \dimen1=\wd1          
        \ifdim\dimen0>\dimen1                   
        \rlap{\hbox to \dimen0{\hfil/\hfil}}    
        #1                                      
        \else                                   
        \rlap{\hbox to \dimen1{\hfil$#1$\hfil}} 
        /                                       
        \fi}                                    %

\begin{document}

\title{Non-collinearity in di-jet fragmentation in electron-positron scattering}

\author{P.J.~Mulders
} 
\affiliation{Theory Group, Nikhef, 1098 XG Amsterdam, the Netherlands}
\affiliation{Department of Physics and Astronomy, Faculty of Science,
VU, 1081 HV Amsterdam, the Netherlands}
\author{C. Van Hulse
}
\email{p.j.g.mulders@vu.nl, cvanhuls@mail.cern.ch}
\affiliation{Department of Theoretical Physics and History of Science, University of the Basque Country UPV/EHU, 48080 Bilbao, Spain}
\affiliation{School of Physics, University College Dublin, Dublin 4, Ireland}
\begin{abstract}
We study fragmentation in electron-positron annihilation assuming a di-jet situation, using variables defined independent of any frame.  In a collinear situation some of the variables are centered around zero with the small deviations attributed to intrinsic transverse momenta and large deviations attributed to additional hard subprocesses. Of course there is a gradual transition. Our modest goal is to show that covariantly defined variables are well suited to get a feeling for the magnitude of intrinsic transverse momenta.
\end{abstract}

\maketitle


\section{Introduction}

\begin{figure}\centering
\epsfig{file=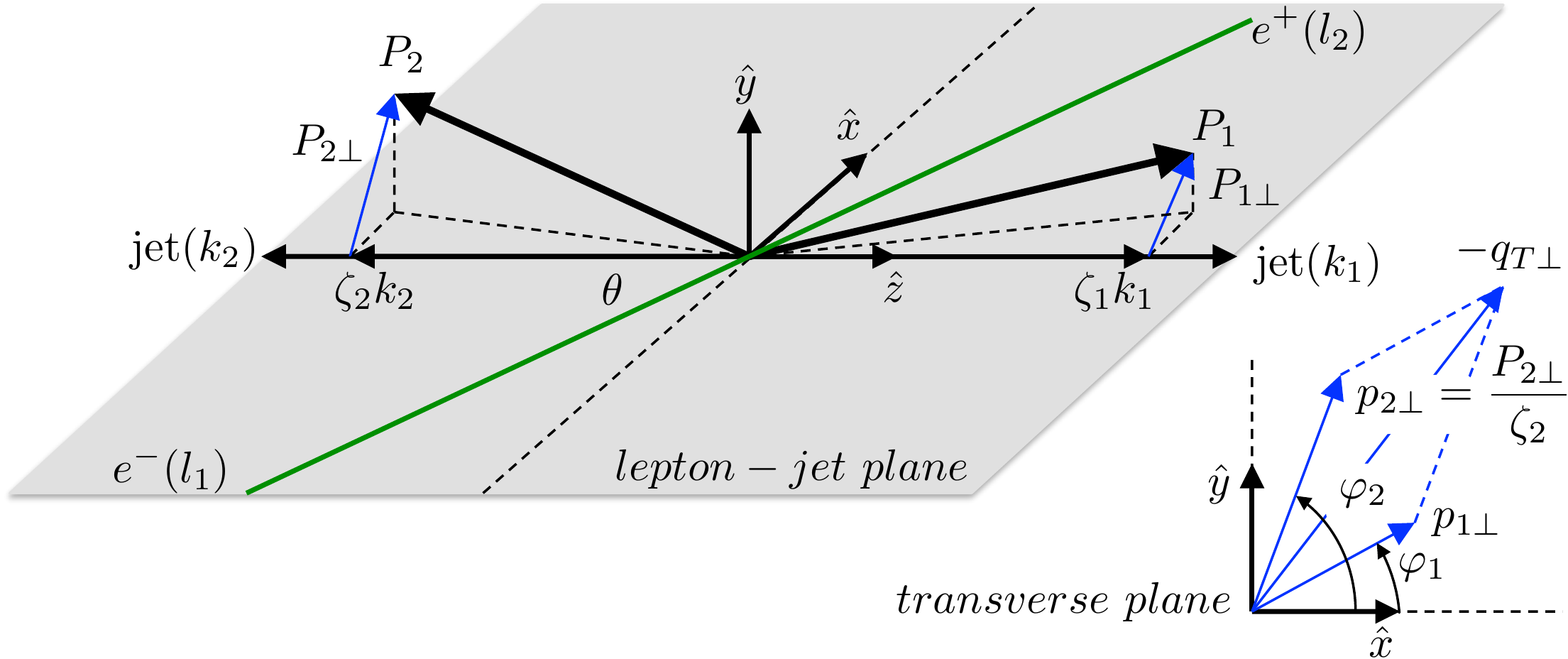,width=0.48\textwidth}
\caption{\label{eeqq12} Basic vectors in $e^-e^+$ annihilation (momenta $l_1$ and $l_2$) into two hadrons with momenta $P_{1}$ and $P_{2}$ originating from two (non-observed) partons with momenta $k_1$ and $k_2$.}
\end{figure}

Inclusive production of two hadrons in electron-positron annihilation is a way to study intrinsic transverse momenta in the fragmentation functions of those hadrons. The non-collinearity in the process can be defined as a spacelike vector $q_\st$ to be specified below. In the basic transverse momentum dependent (TMD) treatment, the $q_\st$ dependence is a folding of intrinsic transverse momenta of the two fragmentation functions in the case of inclusive two-hadron production. Of course, we have in mind the situation shown in figure~\ref{eeqq12} where the two hadrons emerge in two jets with opposite momenta, where in zeroth order the jet directions can be identified with the parent partons. In this note, we just want to look at how scaling variables constructed with invariants in the scattering process can be linked to the hadron momentum fractions in a di-jet situation. One has to keep in mind that partonic transverse momenta are not unique as they depend on a Sudakov expansion for the momenta involving a light-like vector $n$ ($n^2 = 0$) conjugate to the hadron momentum $P\cd n = 0$. The light-like vector acquires a meaning if in a hard process transverse momenta are probed and it then depends on a combination of hard momenta in the scattering process. This is relevant for mass corrections (from hadron masses) and factorisation. In this note, we are just after observables that like rapidity or pseudo-rapidity are useful as measures of non-collinearity and intrinsic transverse momenta.
  
\section{Basic definitions}

\begin{figure*}\centering
\epsfig{file=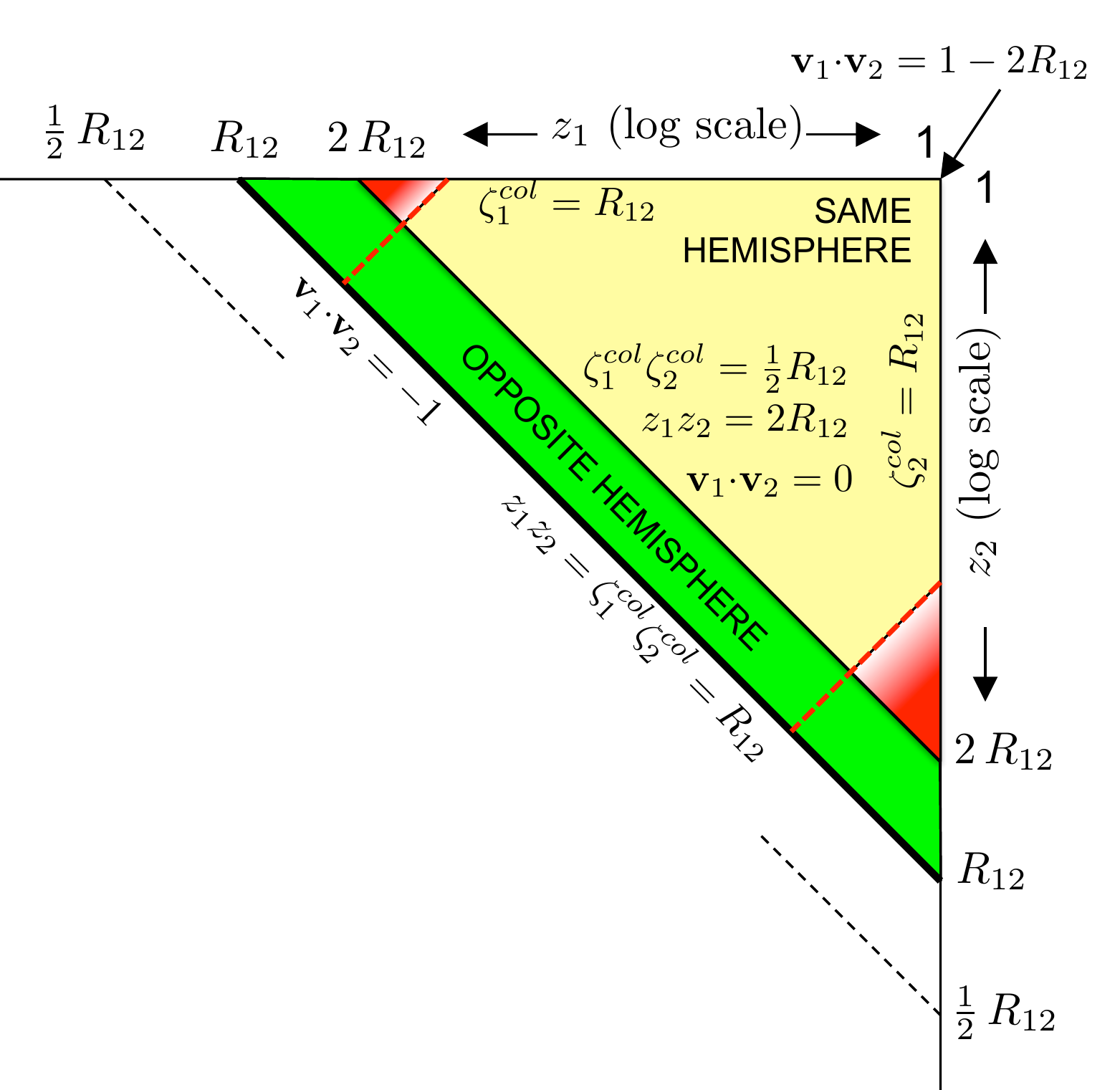,width=0.4\textwidth}
\hspace{1cm}
\epsfig{file=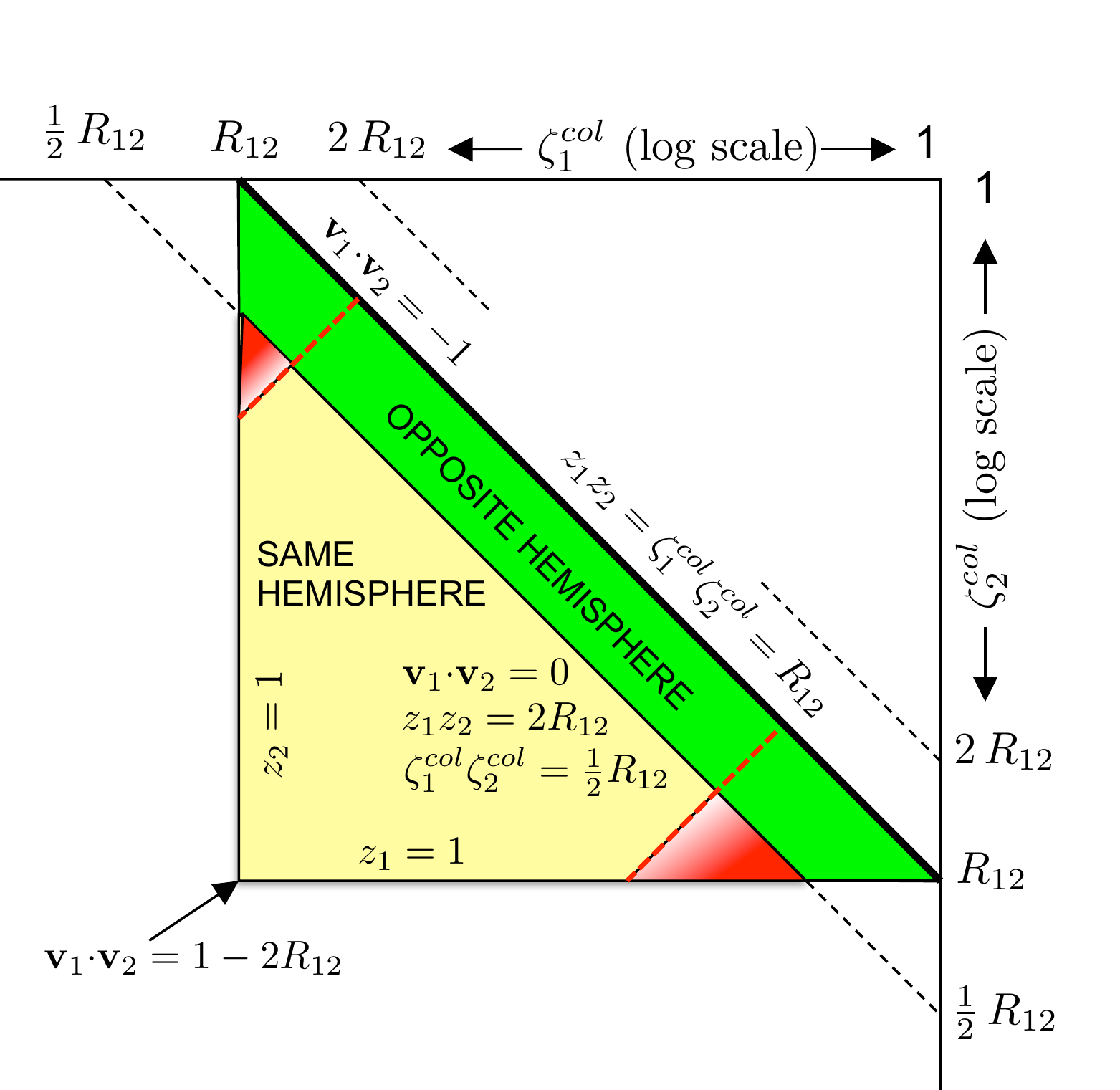,width=0.4\textwidth}
\caption{\label{fig:range} Range of $z_i$ and $\zeta_i^{col}$ for a given value of the ratio $R_{12} = 2\,P_1{\cdot}P_2/s$, divided in opposite and same hemisphere events according to Eq.~\ref{eq:pp_zz_cut}. An indication of the regions where (for sufficiently small $R_{12}$) the quantities $\zeta_i$ turn negative is indicated as the red triangular regions (see discussion following Eq.~\ref{eq:zeta-relations}). The distance of points to the line $z_1z_2=\zeta_1^{col}\zeta_2^{col} = R_{12}$ is a measure of $q_\st^2/s$ (see Eqs.~\ref{eq:qT2} and \ref{eq:z_zprim_approx}).}
\end{figure*}

We look at $e^-(l_1) + e^+ (l_2) \rightarrow H_1(P_{1}) + H_2(P_{2}) + \mbox{anything}$, having in mind an intermediate step into two partons $e^-(l_1) + e^+ (l_2) \rightarrow \mbox{parton}_1 (k_1) + \mbox{parton}_2(k_2)$. The di-jet situation is then the case that $k_1^2 = k_2^2 = 0$.
The measure for non-collinearity~\cite{{Ralston:1979ys},Collins:1981uk,Collins:1981uw,Boer:1997mf} is given by
\be 
q_\st \equiv q - \frac{P_{1}}{\zeta_1} - \frac{P_{2}}{\zeta_2},
\label{eq:qTdef}
\ee
with $q = l_1 + l_2 = k_1 + k_2$ and the requirements that $P_{1}\cd q_\st = P_{2}\cd q_\st = 0$ implying that the hadrons determine the light-like directions in which case the transverse directions for two-particle inclusive (2PI) annihilation are given by $g^{\mu\nu}_\st = g^{\mu\nu} - P_{1}^{\{\mu}P_{2}^{\nu\}}/2P_{1}\cd P_{2}$. This is sufficient to solve for what we will refer to as 2PI momentum fractions $\zeta_{1,2}$, which do include mass corrections,
\bea
&&
\zeta_1 = \frac{P_{1}\cd P_{2} - \frac{M_{1}^2 M_{2}^2}{P_{1}\cd P_{2}}}{P_{2}\cd q - \frac{M_{2}^2\,P_{1}\cd q}{P_{1}\cd P_{2}}}, \label{eq:def_zeta1}
\\\mbox{and}\nonumber &&
\\ &&
\zeta_2 = \frac{P_{1}\cd P_{2} - \frac{M_{1}^2 M_{2}^2}{P_{1}\cd P_{2}}}{P_{1}\cd q - \frac{M_{1}^2\,P_{2}\cd q}{P_{1}\cd P_{2}}},
\label{eq:def_zeta2}
\eea
with $M_{i}$ representing the hadron masses.

The (negative) invariant length of $q_\st^2$ signalling non-collinearity can be expressed in the invariants as 
\bea 
q_\st^2 &=& q^2\left(1 - \frac{P_{1}\cd q}{\zeta_1\,q^2} - \frac{P_{2}\cd q}{\zeta_2\,q^2}\right)
\\ &=& q^2\left(1-\frac{2P_{1}\cd P_{2}}{\zeta_1\zeta_2\,q^2}- \frac{M_{1}^2}{\zeta_1^2\,q^2} - \frac{M_{2}^2}{\zeta_2^2\,q^2}\right)
\\&=& \frac{q^2}{2}\left(2 - \frac{z_1}{\zeta_1} - \frac{z_2}{\zeta_2}\right),
\label{eq:qT2}
\eea
where $z_1$ and $z_2$ are the equivalents of the Bjorken scaling variable in deep-inelastic scattering~\cite{Altarelli:1979kv}, 
\be 
z_1 = \frac{2P_{1}\cd q}{q^2} \quad \mbox{and}\quad
z_2 = \frac{2P_{2}\cd q}{q^2}.
\ee
In the $e^-e^+$ center-of-mass frame, henceforth called CM frame, 
these are $z_i = 2E_{i}/\sqrt{s}$, where $s = 2\,l_1\cd l_2$. In the 2PI case, these variables are restricted to $0\le z_i\le 1$~\cite{Altarelli:1979kv}. The 2PI variables $\zeta_1$ and $\zeta_2$ are not constrained to this interval, as will be illustrated further.

The variables $z_i$ together with the ratio
\be
R_{12} = \frac{2\,P_1{\cdot}P_2}{q^2} = \frac{2\,P_1{\cdot}P_2}{s}
\label{eq:defR12}
\ee
can help to identify hadrons that fit a description as belonging to opposite or same parton jets. We note first of all that this ratio satisfies $R_{12} < 1$, with the upper limit, up to mass corrections, reached in an elastic process like $e^+e^- \rightarrow \pi^+\pi^-$. An appropriate lower limit for our considerations would be something like back-to-back pions with a minimal energy $E$ of around $0.5$ GeV in the CM frame, corresponding to $R_{12} \gtrsim$ (1 GeV$^2$)/$s$. In the CM frame the following ratio holds for any pair,
\be
\frac{4\,P_{1} \cdot P_{2}}{z_1z_2\,s}=\frac{E_{1} E_{2}-\bm{P_{1}}\cd\bm{P_{2}}}{E_{1} E_{2}} = 1 - \bm v_{1}\cd\bm v_{2},
\label{eq:pp_zz}
\ee
thus 
\be 
z_1z_2 = \frac{2\,R_{12}}{1-\bm v_{1}\cd\bm v_{2}},
\ee
where the three-momentum scalar product $\bm v_{1}\cd\bm v_{2}$, here and in the following evaluated in the CM frame, is positive for fast hadrons within a cone around the same fragmenting parton, while it is negative for fast hadrons from different parton jets, as indicated in figure~\ref{eeqq12}. 
The limit between both situations, with two fast hadrons perpendicular to each other, 
corresponds to $z_1z_2 = 2\,R_{12}$, resulting in the criterion
\bea
&&
\mbox{opposite hemisphere:}\ R_{12} \le z_1z_2 \le 2\,R_{12},
\nonumber \\&&
\mbox{same hemisphere:}\quad 2\,R_{12} \le z_1z_2 \le 1,
\label{eq:pp_zz_cut}
\eea
as a first-order requirement to characterise hadrons as belonging to back-to-back parton jets or to the same parton jet, respectively.

We want to further illustrate this by a comparison with the collinear fractions $\zeta_1^{col}$ and $\zeta_2^{col}$, scaling variables defined as~\cite{Feynman:1973xc}
\be 
\zeta_1^{col} = \frac{P_{1}\cd P_{2}}{P_{2}\cd q} \quad \mbox{and}\quad
\zeta_2^{col} = \frac{P_{1}\cd P_{2}}{P_{1}\cd q}.
\ee
These variables are also limited to $0\leq \zeta_i^{col}\leq 1$~\cite{Altarelli:1979kv}.
We note that
\be 
z_1 \zeta_2^{col} = \zeta_1^{col}z_2 = R_{12}.
\label{eq:zzcol_rel}
\ee
The range of the scaling variables $\zeta_i^{col}$ and $z_i$ is illustrated in figure~\ref{fig:range}. 
Using $\epsilon_i = M_{i}^2/P_{1}\cd P_{2}$ we have the exact relations
\be 
\zeta_1 = \zeta_1^{col}\frac{1-\epsilon_1\epsilon_2}{1-\epsilon_2\,\zeta_1^{col}/\zeta_2^{col}},\ 
\zeta_2 = \zeta_2^{col}\frac{1-\epsilon_1\epsilon_2}{1-\epsilon_1\,\zeta_2^{col}/\zeta_1^{col}}.
\label{eq:zeta-relations}
\ee
From it, we obtain the region where $\zeta$ becomes negative, e.g.,~$\zeta_1 < 0$, when $\zeta_2^{col} < \epsilon_2\,\zeta_1^{col} = (2M_2^2/R_{12}s)\zeta_1^{col}$, or $z_2 < (2M_2^2/R_{12}s)z_1$. It just signals that one is outside the region of validity for simple parton branching and fragmentation. Note that the regions for hadrons 1 and 2 may be different (hence the asymmetry of the effect in figure~\ref{fig:range}).
Up to $\epsilon^2$ terms we have (if $\zeta_2^{col}$ is not too small) 
\bea
&&
\zeta_1 \approx \zeta_1^{col}\left(1 + \epsilon_2 \zeta_1^{col}/\zeta_2^{col}\right),
\eea
(thus $\zeta_i \gtrsim \zeta_i^{col}$) and for sufficiently small $\epsilon_i$, for instance for pions,
\bea
\zeta_{1/2} \approx z_{1/2}\left( 1 + \frac{q_\st^2}{q^2}\right) 
\label{eq:z_zprim_approx}
\eea
(thus $\zeta_i < z_i$ for small $\epsilon_i$). 

There is a second measure for non-collinearities, which is the determinant of the four measured four-momenta,
\be
D_\st \equiv -\frac{4\,\epsilon^{l_1l_2P_{1}P_{2}}}{\zeta_1\zeta_2\,s^{3/2}}= -\frac{4\,\epsilon_{\mu\nu\rho\sigma}l_1^\mu l_2^\nu P_{1}^{\rho} P_{2}^\sigma}{\zeta_1\zeta_2\,s^{3/2}}.
\label{eq:DT_def}
\ee
Here, a particular normalisation has been chosen, a choice that will become clear in the next section via the link to the intrinsic 
transverse momentum. The invariant $q_\st^2$ only signals that the lepton momenta and hadron momenta are not in a single plane, 
while the determinant contains also information on the relative orientations of the hadrons.

\section{Interpreting non-collinearities}

For the interpretation of the non-collinearities defined in the previous section, it is convenient to make the four vectors explicit in the leptonic and partonic restframe, using either $p = (p^0,p^1,p^2,p^3)$ or $p = [p^-,p^+,p_\st^1,p_\st^2]$ with $p^\pm = (p^0\pm p^3)/\sqrt{2}$, and starting with
\bea
&&l_1 = \frac{\sqrt{s}}{2} \left(\begin{array}{c} 1 \\ 0 \\ 0 \\ 1 \end{array}\right)
= \sqrt{\frac{s}{2}}\left[\begin{array}{c} 0 \\ 1 \\ 0 \\ 0 \end{array}\right],
\nonumber \\ &&
l_2 = \frac{\sqrt{s}}{2} \left(\begin{array}{c} 1 \\ 0 \\ 0 \\ -1 \end{array}\right)
= \sqrt{\frac{s}{2}}\left[\begin{array}{c} 1 \\ 0 \\ 0 \\ 0 \end{array}\right],
\eea
and the partonic light-like momenta 
\bea
&&
k_1 = \frac{\sqrt{s}}{2} \left(\begin{array}{c}1 \\ \sin\theta \\ 0 \\ \cos\theta \end{array}\right)
= \sqrt{\frac{s}{2}}\left[\begin{array}{c}
- t/s \\  - u/s \\  \sqrt{2tu}/s \\  0
\end{array}\right],
\nonumber \\ &&
k_2 = \frac{\sqrt{s}}{2} \left(\begin{array}{c}1 \\ -\sin\theta \\ 0 \\ -\cos\theta \end{array}\right) 
= \sqrt{\frac{s}{2}}\left[\begin{array}{c}
- u/s \\  - t/s \\  -\sqrt{2tu}/s \\  0
\end{array}\right].
\eea
We have used here the Mandelstam variables in the process $e^-e^+ \rightarrow \mbox{jet}_1 \mbox{jet}_2$, 
\begin{eqnarray}
s & = & q^2 = 2\,l_1\cd l_2,
\nonumber\\
 -t & = & 2\,l_1\cd k_1 
= s\,\sin^2\left(\tfrac{1}{2}\theta\right)\quad\mbox{}
\nonumber \\
-u & = & 2\,l_1\cd k_2 
= s\,\cos^2\left(\tfrac{1}{2}\theta\right)
\end{eqnarray}
and angle $\theta$ for which
\begin{equation}
\cos\theta = \frac{t-u}{s}, \quad \sin\theta = \frac{2\sqrt{tu}}{s}.
\end{equation}
If we take the jet direction as z-axis, we are in a frame in which the momenta of the hadrons will have perpendicular components indicated as pseudo-quark momenta $p_{i\perp} \equiv P_{i\perp}/\zeta_i$. The hadron momenta can be written as
\bea
&&\frac{P_{1}}{\zeta_1} 
= \frac{\sqrt{s}}{2}\left(\begin{array}{c}
1 + \mu_{1\perp}^2/s \\ 
2\,p_{1\perp}^x/\sqrt{s} \\ 2\,p_{1\perp}^y/\sqrt{s}
\\ 1 - \mu_{1\perp}^2/s
\end{array}\right)
= \sqrt{\frac{s}{2}}\left[\begin{array}{c}
\mu_{1\perp}^2/s \\ 1 \\ 
p_{1\perp}^x\sqrt{2/s} \\ p_{1\perp}^y\sqrt{2/s}
\end{array}\right], 
\nonumber\\&&
\frac{P_{2}}{\zeta_2}  
= \frac{\sqrt{s}}{2}\left(\begin{array}{c}
1 + \mu_{2\perp}^2/s \\ 
2\,p_{2\perp}^x/\sqrt{s} \\ 2\,p_{2\perp}^y/\sqrt{s}
\\ -1 + \mu_{2\perp}^2/s
\end{array}\right)
= \sqrt{\frac{s}{2}}\left[\begin{array}{c}
1 \\ \mu_{2\perp}^2/s \\ 
p_{2\perp}^x\sqrt{2/s} \\ p_{2\perp}^y\sqrt{2/s}
\end{array}\right],
\nonumber \\&& 
\label{hadron-momenta}
\eea
where $\mu^2_{i\perp} \equiv (M_{i}^2 - P_{i\perp}^2)/\zeta_i^2 = M_{i}^2/\zeta_i^2 + \bm p_{i\perp}^2$. These momenta are measurable quantities if the jet axis (e.g.~identified with a thrust axis) is known, since the angles assume the lepton-jet plane to be the x-z plane. In this frame the partonic momenta do not have transverse components, thus $k_{i\perp} \equiv 0$. Thus explicitly the partonic momenta are
\bea
&&k_1 = \frac{\sqrt{s}}{2} \left(\begin{array}{c} 1 \\ 0 \\ 0 \\ 1 \end{array}\right)
= \sqrt{\frac{s}{2}}\left[\begin{array}{c} 0 \\ 1 \\ 0 \\ 0 \end{array}\right],
\nonumber \\ &&
k_2 = \frac{\sqrt{s}}{2} \left(\begin{array}{c} 1 \\ 0 \\ 0 \\ -1 \end{array}\right)
= \sqrt{\frac{s}{2}}\left[\begin{array}{c} 1 \\ 0 \\ 0 \\ 0 \end{array}\right],
\eea
while choosing them along the z-axis implies that the vectors in Eq.~\ref{hadron-momenta} are rotated over the angle $\theta$ in the x-z plane, 
and
\bea
&&
l_1 = \frac{\sqrt{s}}{2} \left(\begin{array}{c}1 \\ \sin\theta \\ 0 \\ \cos\theta \end{array}\right)
= \sqrt{\frac{s}{2}}\left[\begin{array}{c}
- t/s \\  - u/s \\  \sqrt{2tu}/s \\  0
\end{array}\right],
\nonumber \\ &&
l_2 = \frac{\sqrt{s}}{2}\left(\begin{array}{c}1 \\ -\sin\theta \\ 0 \\ -\cos\theta \end{array}\right) 
= \sqrt{\frac{s}{2}}\left[\begin{array}{c}
- u/s \\  - t/s \\  -\sqrt{2tu}/s \\  0
\end{array}\right].
\eea
Rotating the lepton momenta is less cumbersome than rotating the hadron momenta in Eqs.~\ref{hadron-momenta}.

For the jet momenta, which are just light-like directions, we have
\bea
&&
k_1 = \frac{P_{1}}{\zeta_1} + p_{1\perp} - \frac{\mu_{1\perp}^2}{s}\,k_2,
\label{eq:k1jet}
\\&&
k_2 = \frac{P_{2}}{\zeta_2} + p_{2\perp} - \frac{\mu_{2\perp}^2}{s}\,k_1.
\label{eq:k2jet}
\eea
We nicely have $\zeta_1 = P_{1}\cd k_2/k_1\cd k_2$ and $\zeta_2 = P_{2}\cd k_1/k_1\cd k_2$. The non-collinearity measures introduced in the previous section are
\bea
q_\st &=& -p_{1\perp} - p_{2\perp} - \frac{\mu_{2\perp}^2}{s}k_1 - \frac{\mu_{1\perp}^2}{s}k_2,
\\
q_\st^2 &=& (p_{1\perp} + p_{2\perp})^2 + \frac{\mu_{1\perp}^2\mu_{2\perp}^2}{s}
\nonumber\\&\approx& (p_{1\perp} + p_{2\perp})^2
\label{qT2-1}
\eea
and with $p_{i\perp} = (p_{i\perp}^x,p_{i\perp}^y) = \vert p_{i\perp}\vert (\cos\varphi_i, \sin\varphi_i)$, we get
\bea
D_\st &=& -\vert p_{1\perp}\vert\sin\theta \,\sin\varphi_1\left(1-\frac{\mu_{2\perp}^2}{s}\right)
\nonumber \\&&
\mbox{} - \vert p_{2\perp}\vert\sin\theta \,\sin\varphi_2\left(1-\frac{\mu_{1\perp}^2}{s}\right)
\nonumber \\ && \mbox{}-\frac{\vert p_{1\perp}\vert\,\vert p_{2\perp}\vert}{\sqrt{s}} \cos\theta\,\sin(\varphi_1-\varphi_2)
\nonumber \\
&\approx & -(p_{1\perp}^y + p_{2\perp}^y)\sin\theta
+\frac{(p_{1\perp}^x\, p_{2\perp}^y - p_{1\perp}^y\,p_{2\perp}^x)}{\sqrt{s}} \cos\theta
\nonumber\\&\approx & q_\st^y\sin\theta + \frac{\bm p^{cm}_{1\perp}\times \bm p^{cm}_{2\perp}}{\sqrt{s}}\,\cos\theta,
\label{DT-1}
\eea
where (illustrated in figure~\ref{eeqq12}) the $y$-direction is the one orthogonal to the plane formed by the lepton and (back-to-back) jet directions.
In Eq.~\ref{DT-1} one recognises
\be
\epsilon^{l_1P_1l_2P_2}/\zeta_1\zeta_2 \approx \epsilon^{l_1l_2k_1p_{2\perp}}+ \epsilon^{l_1l_2p_{1\perp}k_2} - \epsilon^{l_1l_2p_{1\perp}p_{2\perp}},
\label{eq:expandD}
\ee
following from Eqs.~\ref{eq:k1jet} and \ref{eq:k2jet} omitting mass corrections. The third term in Eq.~\ref{eq:expandD} corresponds to the $1/\sqrt{s}$ suppressed term in Eq.~\ref{DT-1}.

In this section, knowledge of all components of the hadrons in the (theoretical) lepton-parton CM frame has been assumed.
The magnitude of the combined transverse-momentum components can be extracted from the lepton and hadron momenta in the lepton CM frame via the approximate
relation in Eq.~\ref{qT2-1}, but the azimuthal orientation of the hadrons cannot be determined as is clear from Eq.~\ref{DT-1}.
It is only possible to obtain information on the azimuthal orientation in the lepton-parton CM frame if we know the partonic axis.  
The latter can be reasonably well approximated by the thrust axis.
In order to study this, let us consider hadron 1 as the one to be studied, thus $1 \rightarrow h$ and identify hadron 2 with jet 2, $k_2 = P_2/\zeta_2$. In that case $M_2 = \vert p_{2\perp}\vert = 0$, the angle $\varphi_2$ becomes undefined and $\zeta_{2}=1$.
We get the information on hadron $1 = h$ obviously from $q_\st = -p_{h\perp} = -P_{h\perp}/\zeta_h$,
\bea
q_\st^2 &=& -\vert p_{h\perp}\vert^2
\label{qT2-2} 
\\
D_\st &=& -\vert p_{h\perp}\vert\sin\theta \,\sin\varphi_h = q_\st^y\sin\theta ,
\label{DT-2}
\eea
everything measured with respect to the jet direction. 

\section{Features in data}

\begin{figure*}\centering
\begin{minipage}{0.48\textwidth}\centering
\epsfig{file=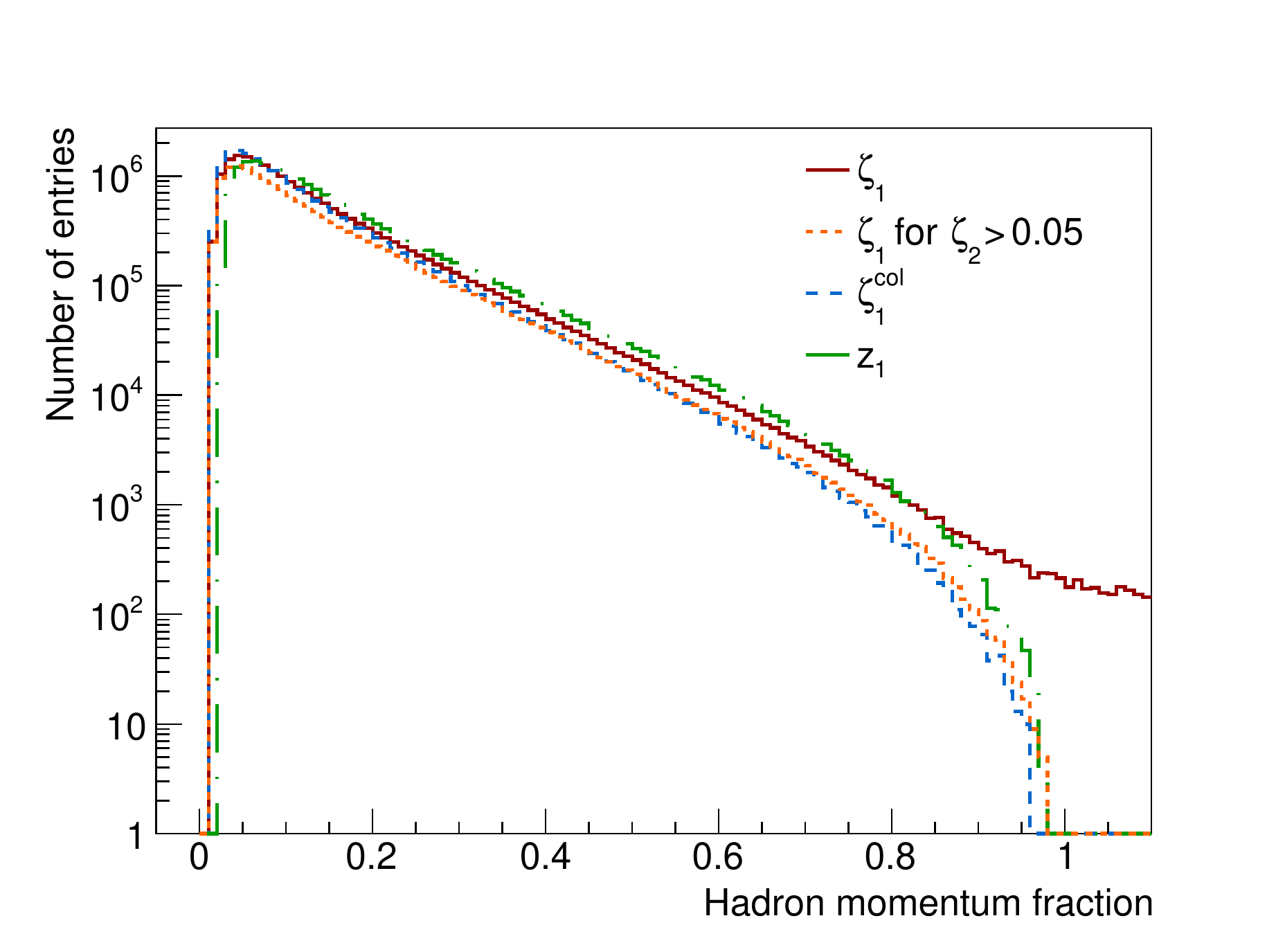,width=1.0\textwidth}
\end{minipage}
\begin{minipage}{0.48\textwidth}\centering
\epsfig{file=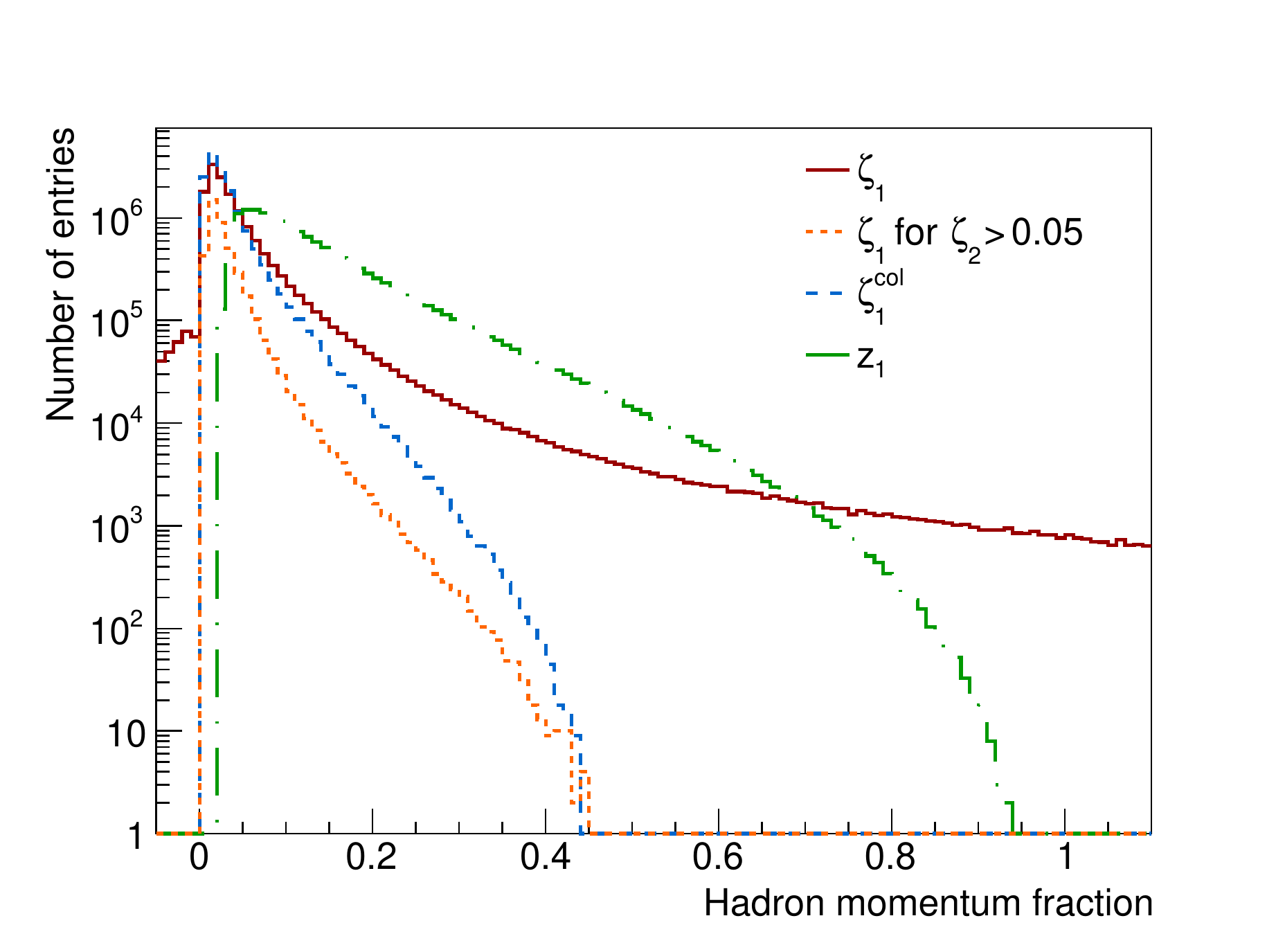,width=1.0\textwidth}
\end{minipage}
\caption{Hadron momentum fractions $\zeta_{1}$ (continuous, red line), $\zeta_{1}$ for $\zeta_{2}>0.05$ (short-dashed, orange line), $\zeta^{col}_1$ (large-dashed, blue line), and $z_{1}$ (dash-dotted, green line) for hadrons in the opposite (left) and same (right) hemisphere, based on their relative orientation.} 
\label{fig:z_rel}
\end{figure*}
 
\begin{figure}[b]\centering
\epsfig{file=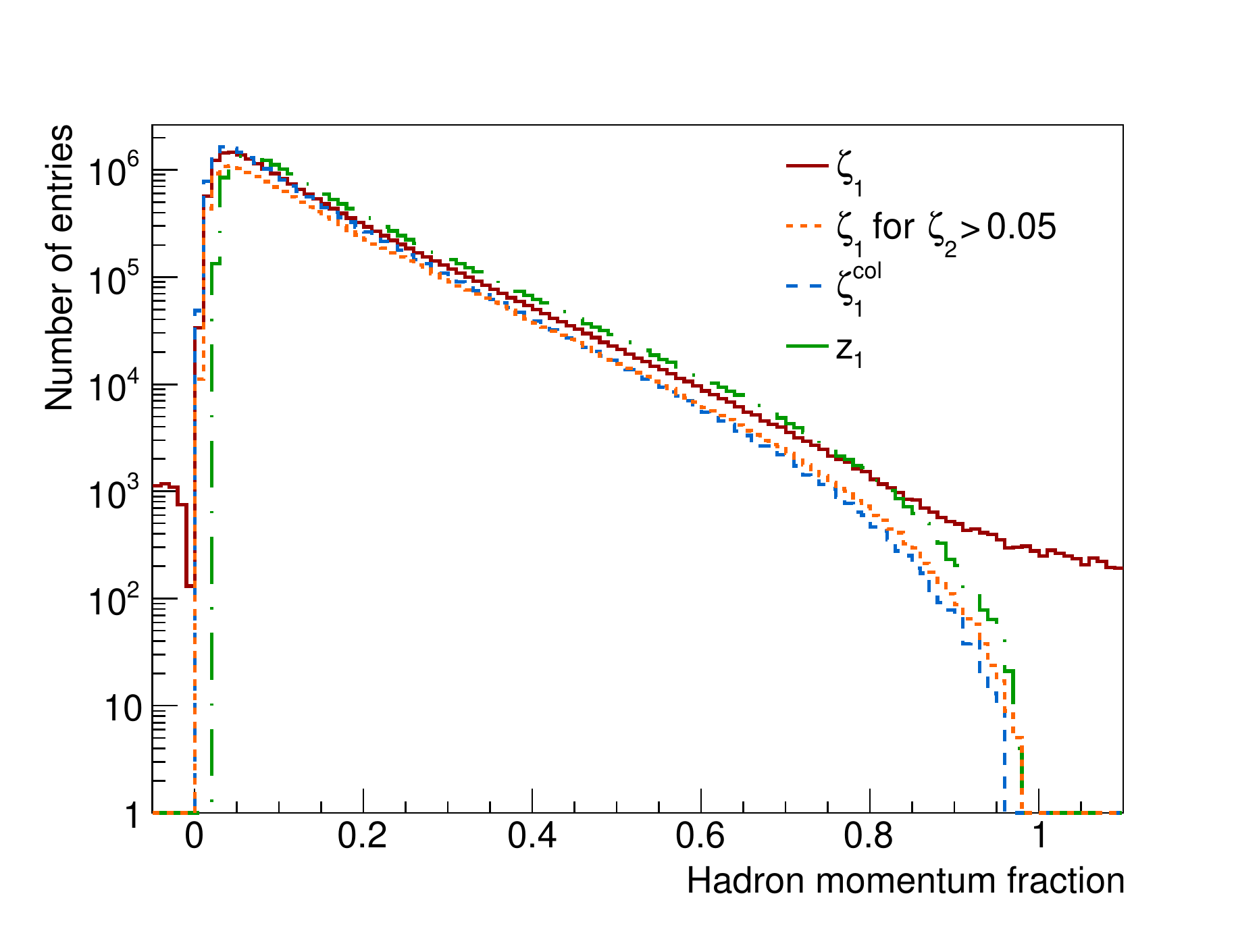,width=0.5\textwidth}
\caption{\label{fig:z_opp} Hadron momentum fractions $\zeta_{1}$ (continuous, red line), $\zeta_{1}$ for $\zeta_{2}>0.05$ (short-dashed, orange line), $\zeta^{col}_1$ (large-dashed, blue line), and $z_{1}$ (dash-dotted, green line) for hadrons in opposite hemispheres, 
based on their orientation with respect to the thrust axis.} 
\end{figure}

\begin{figure}\centering
\epsfig{file=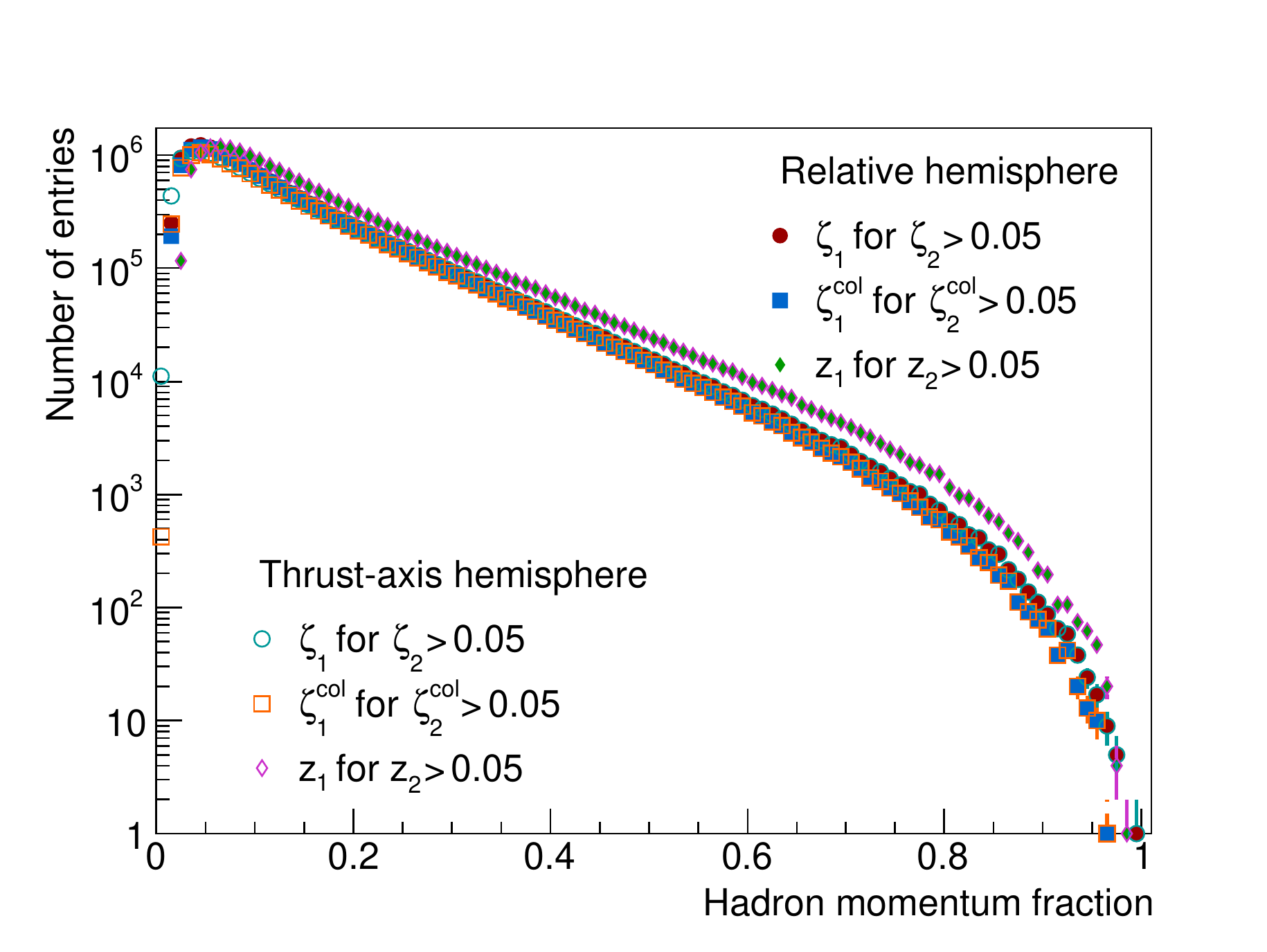,width=0.5\textwidth}
\caption{\label{fig:z_rel_abs_hemi} Hadron momentum fractions $\zeta_{1}$ for $\zeta_{2}>0.05$ (circles), $\zeta^{col}_1$ for $\zeta^{col}_2>0.05$ (squares), 
and $z_{1}$ for $z_{2}>0.05$ (diamonds) for hadrons in opposite hemispheres based on their relative orientation 
(filled symbols) and based on their orientation with respect to the thrust axis (open symbols). 
The vertical error bars represent the square root of the number of entries in each bin.}
\end{figure}

In this section, observables discussed above are illustrated using a PYTHIA Monte-Carlo simulation~\cite{Sjostrand:2000wi} of the process 
$e^+e^-\rightarrow q\bar{q}$ at a CM energy of 10.58~GeV, where $q (\bar{q})$ represents an (anti-)up, down or strange quark. 
This energy corresponds to the $\Upsilon(4S)$ mass, and is characteristic for data collection at the Belle and BaBar experiments~\cite{Bevan:2014iga}. 
Quantum electrodynamics radiative effects are absent in the simulation, and hadron pairs are selected based on the value of their generated momentum. Unless stated otherwise, results are shown for pairs of positively charged pions.

\begin{figure*}\centering
\begin{minipage}{0.48\textwidth}\centering
\epsfig{file=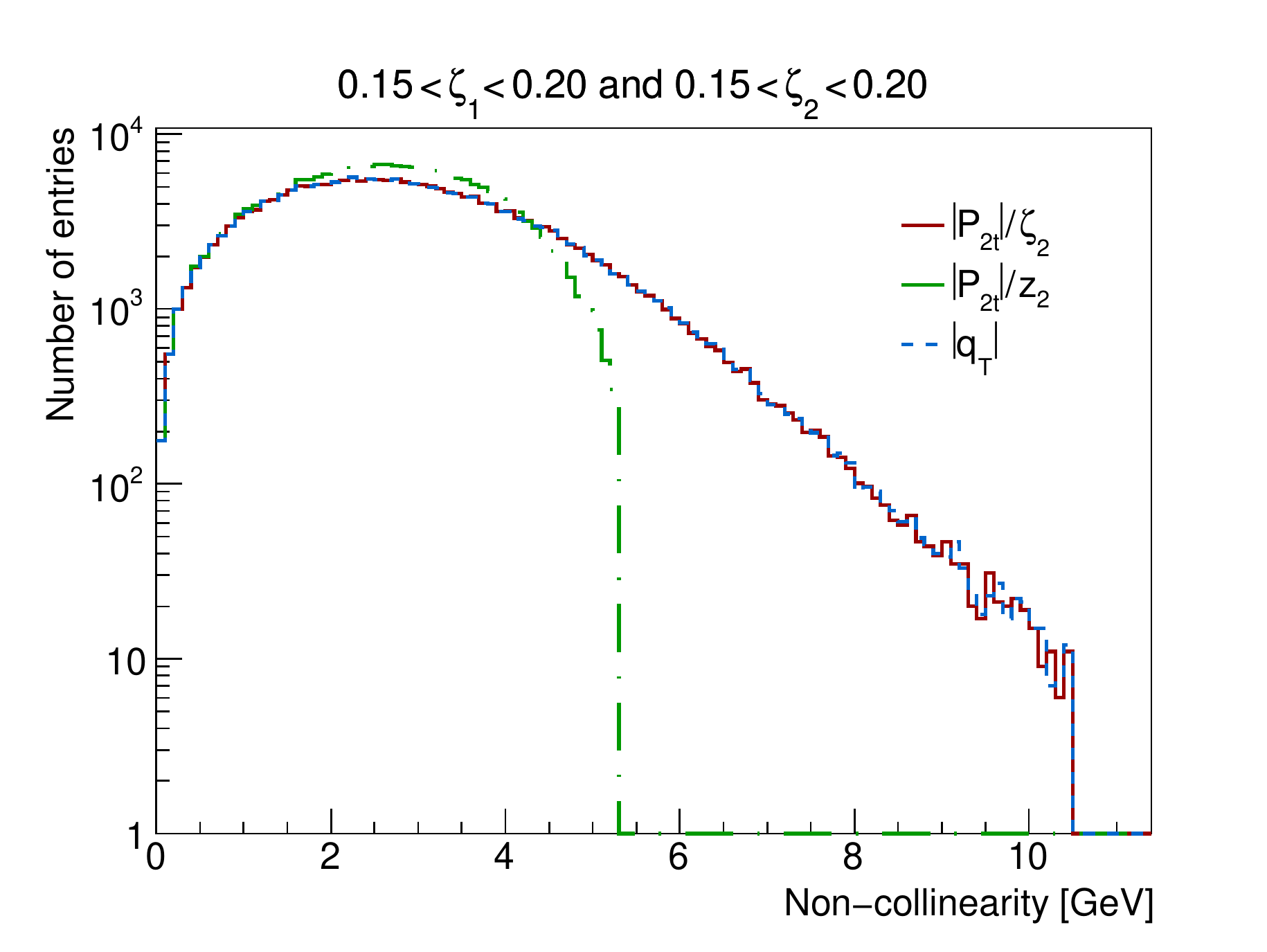,width=1.\textwidth}
\end{minipage}
\begin{minipage}{0.48\textwidth}\centering
\epsfig{file=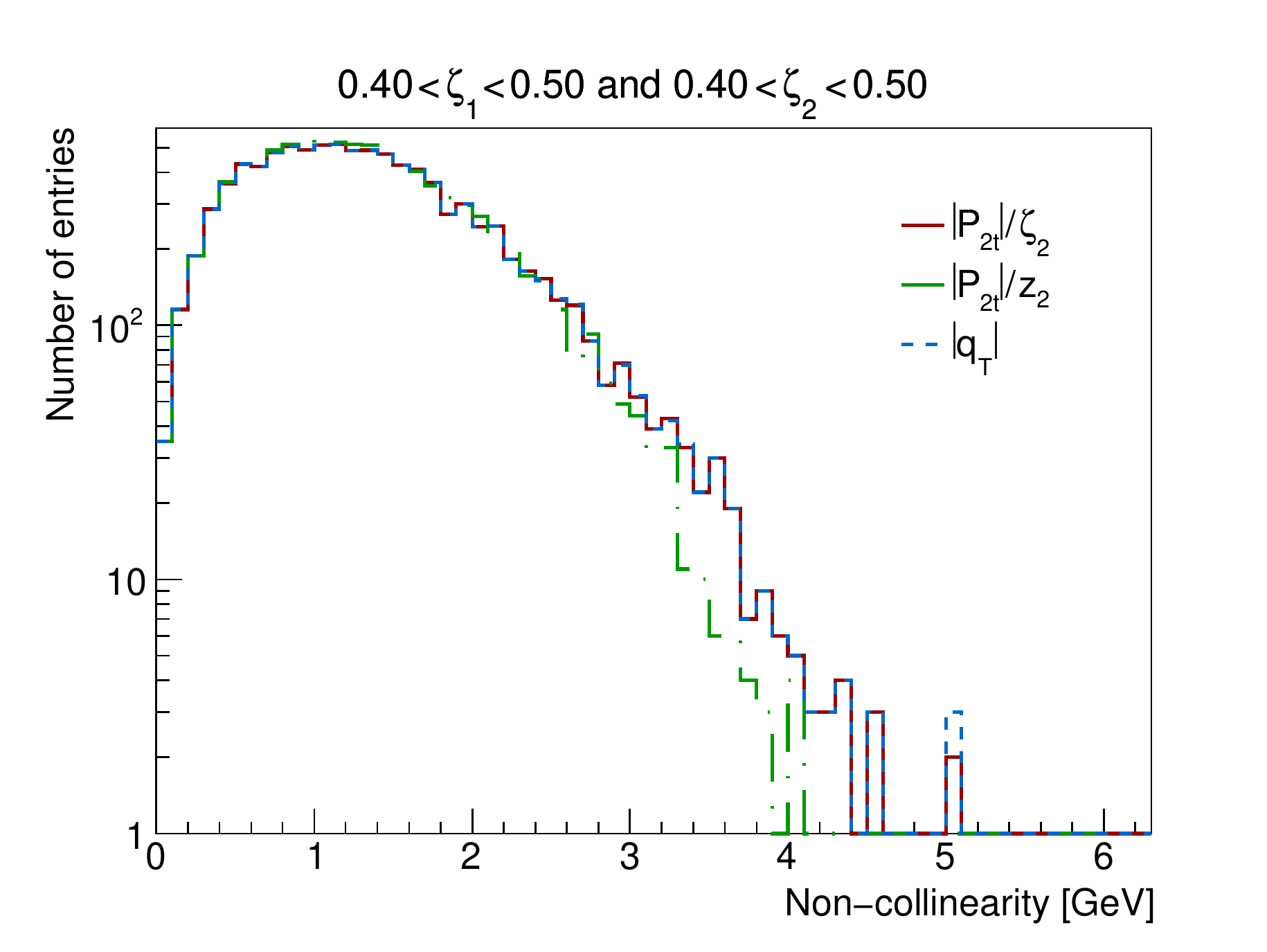,width=1.\textwidth}
\end{minipage}
\caption{The non-collinearity variables $\sqrt{-q_\st^2}$ (dashed, blue line), $\vert P_{2t}\vert/\zeta_{2}$ (continuous, red line), and $\vert P_{2t}\vert/z_{2}$ (dash-dotted, green line) for
$0.15< \zeta_1=\zeta_2 < 0.20$ (left) and $0.40 < \zeta_1=\zeta_2 < 0.50$ (right) for hadrons in opposite hemispheres.}  
\label{fig:ktz}
\end{figure*}

In figure~\ref{fig:z_rel}, left, the hadron momentum fractions $\zeta_{1}$ (continuous, red line), 
$\zeta^{col}_1$ (large-dashed, blue line), and $z_{1}$ (dash-dotted, green line) 
are presented for hadrons that lie in the hemisphere opposite to that 
of the second hadron in the event. In order to fulfil the latter condition, only hadron pairs with negative three-momentum 
scalar product are considered. As can be seen from the figure, the variables $\zeta^{col}_1$ and $z_{1}$ are contained in the interval between 
0 and 1\footnote{In the limit of massless hadrons or infinite CM energy, the values 0 and 1 would effectively be reached.}, 
while the value of $\zeta_{1}$ exceeds 1. If the variable $\zeta_{2}$ of the second
hadron is required to be not too small, not smaller than 0.05 in the present case, 
then $\zeta_{1}$ is also constrained to $0\leq \zeta_{1}\leq 1$, as illustrated by the short-dashed, orange curve, 
and it follows the distribution of $\zeta_1^{col}$.

When selecting hadrons that lie in the same hemisphere, as shown in figure~\ref{fig:z_rel}, right, a larger amount of hadrons 
have a momentum fraction $\zeta_{1}$  larger than 1  (continuous, red line). In addition, also negative values for $\zeta_{1}$ are now observed. 
Requiring again $\zeta_2>0.05$, restricts $\zeta_{1}$ 
to $0.0\leq \zeta_{1}\leq 0.5$ (short-dashed, orange line). 
Also $\zeta_{1}^{col}$ is confined to this interval, while $z_{1}$ lies as 
before between 0 and 1. Hadrons originating from the same quark need to share half of the available CM energy.
This is reflected in the upper limit of $0.5$ for $\zeta_{1}$ and $\zeta_{1}^{col}$.

If the hemisphere selection of hadrons is not based on their relative orientation, but on their orientation with respect to 
the thrust axis, a small fraction of hadrons have a negative $\zeta_{1}$ value when considering the configuration of opposite hemispheres, as can 
be seen in figure~\ref{fig:z_opp} (continuous, red line). These hadrons correspond to pairs that are associated to the same hemisphere
when basing the selection criterion on the hadrons' relative orientation. Requiring $\zeta_{2}>0.05$ removes these hadrons from the selection 
(short-dashed, orange line). 

\begin{figure*}\centering
\begin{minipage}{0.50\textwidth}\centering
\epsfig{file=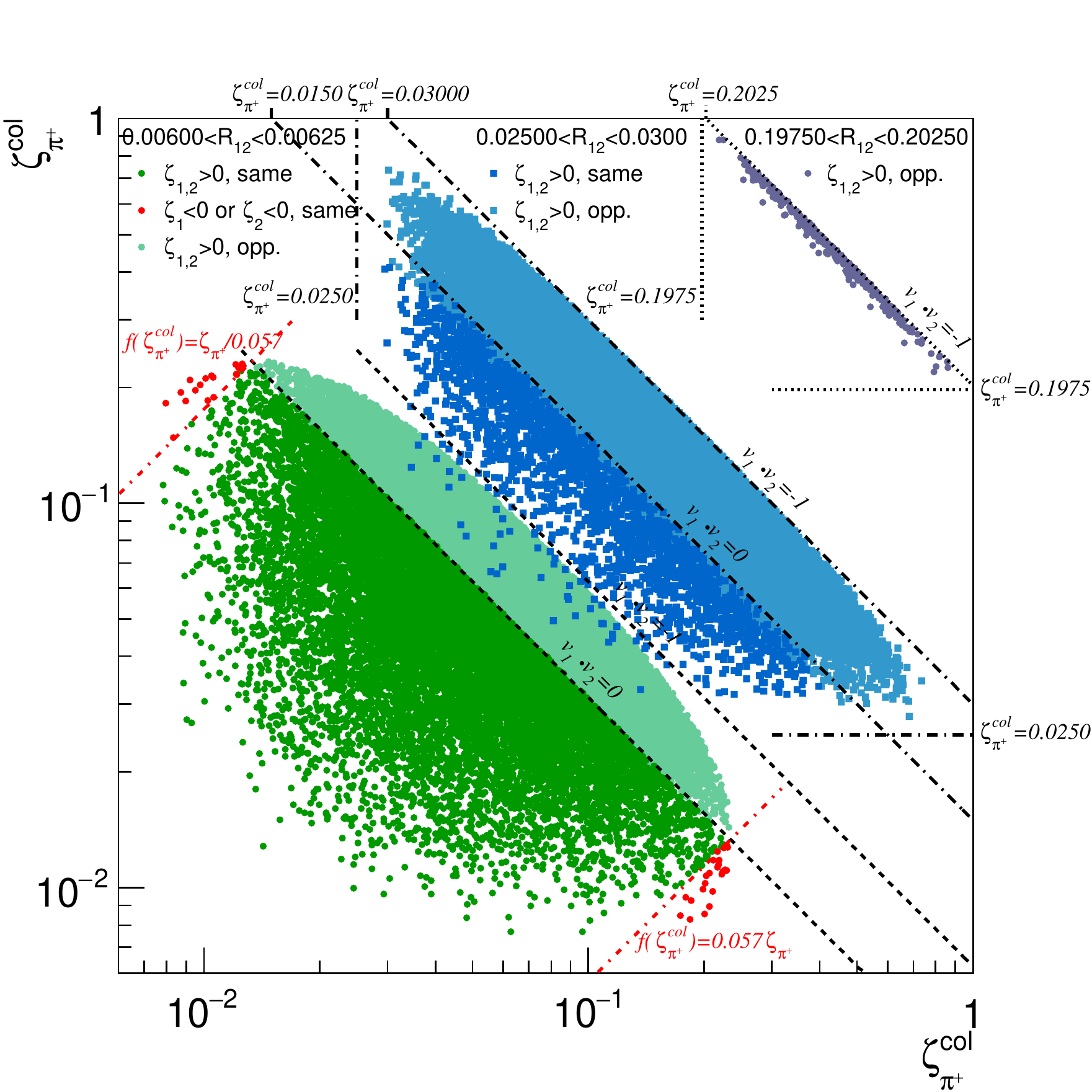,width=1.\textwidth}
\end{minipage}
\begin{minipage}{0.49\textwidth}\centering
\epsfig{file=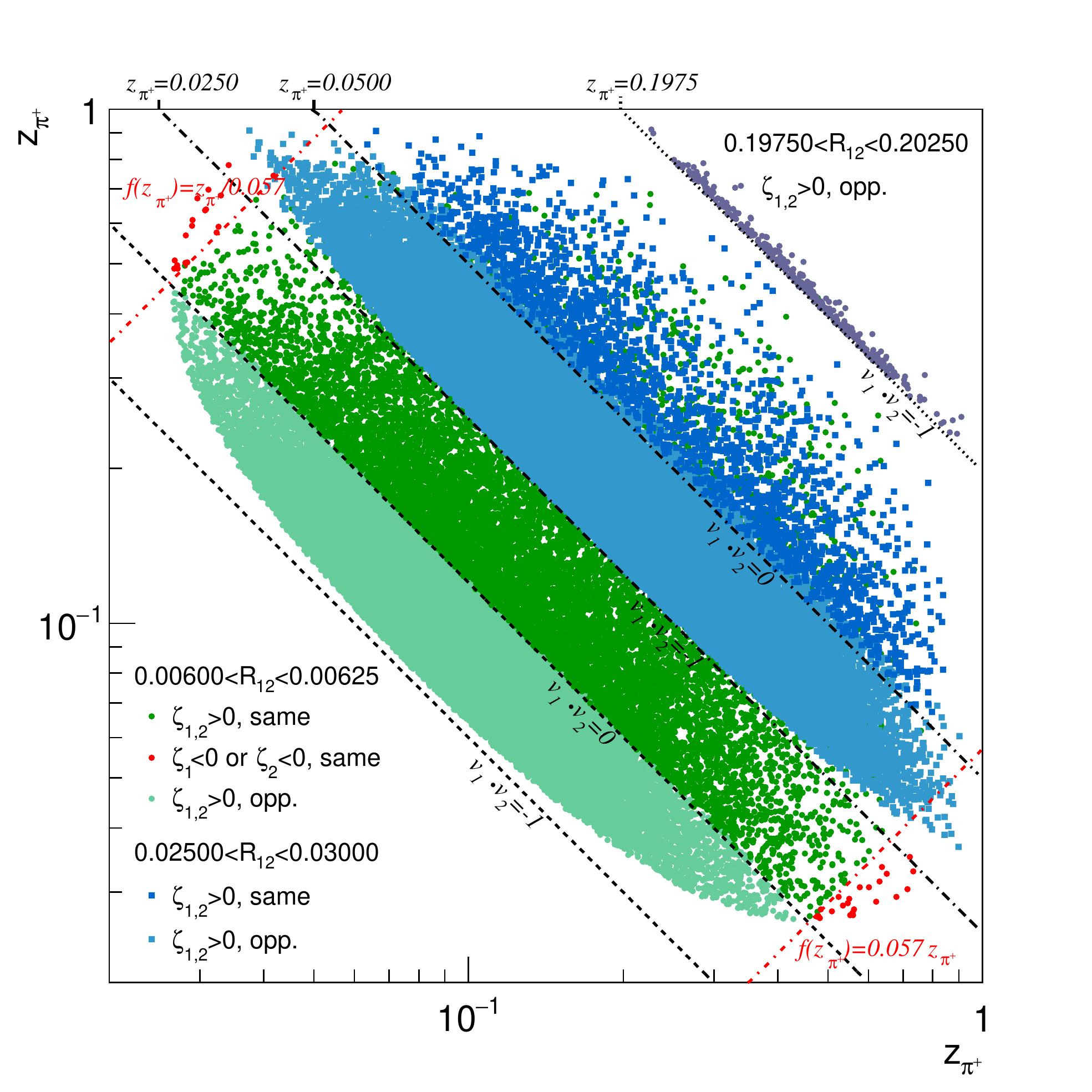,width=1.\textwidth}
\end{minipage}
\caption{Distributions of $\zeta^{col}_2$ versus $\zeta^{col}_1$ (left) and of $z_2$ versus $z_1$ (right) 
for three ranges in $R_{12}$: $0.00600 < R_{12} < 0.00625$, $0.02500 < R_{12} < 0.03000$,  
and $0.19750 < R_{12} < 0.20250$. 
The hemisphere configuration and the sign of $\zeta_i$ are indicated in the figure. 
The diagonal lines indicate the limit for which $v_{1}\cdot v_{2}$ is equal to $0$ or $-1$. 
The vertical and horizontal lines correspond to the lower limit of the respective $R_{12}$ intervals.}
\label{fig:scatter}
\begin{minipage}{0.50\textwidth}\centering
\epsfig{file=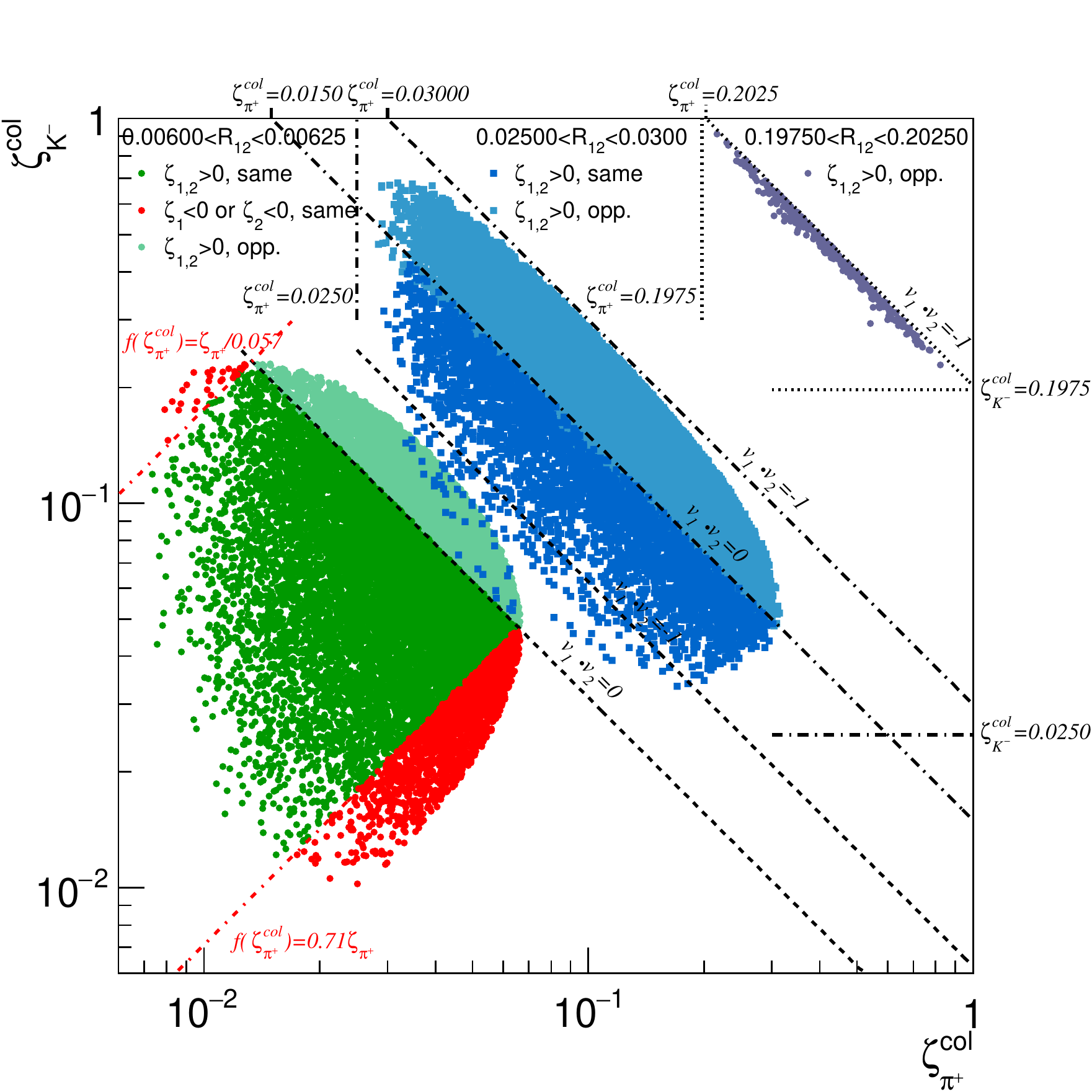,width=1.\textwidth}
\end{minipage}
\begin{minipage}{0.49\textwidth}\centering
\epsfig{file=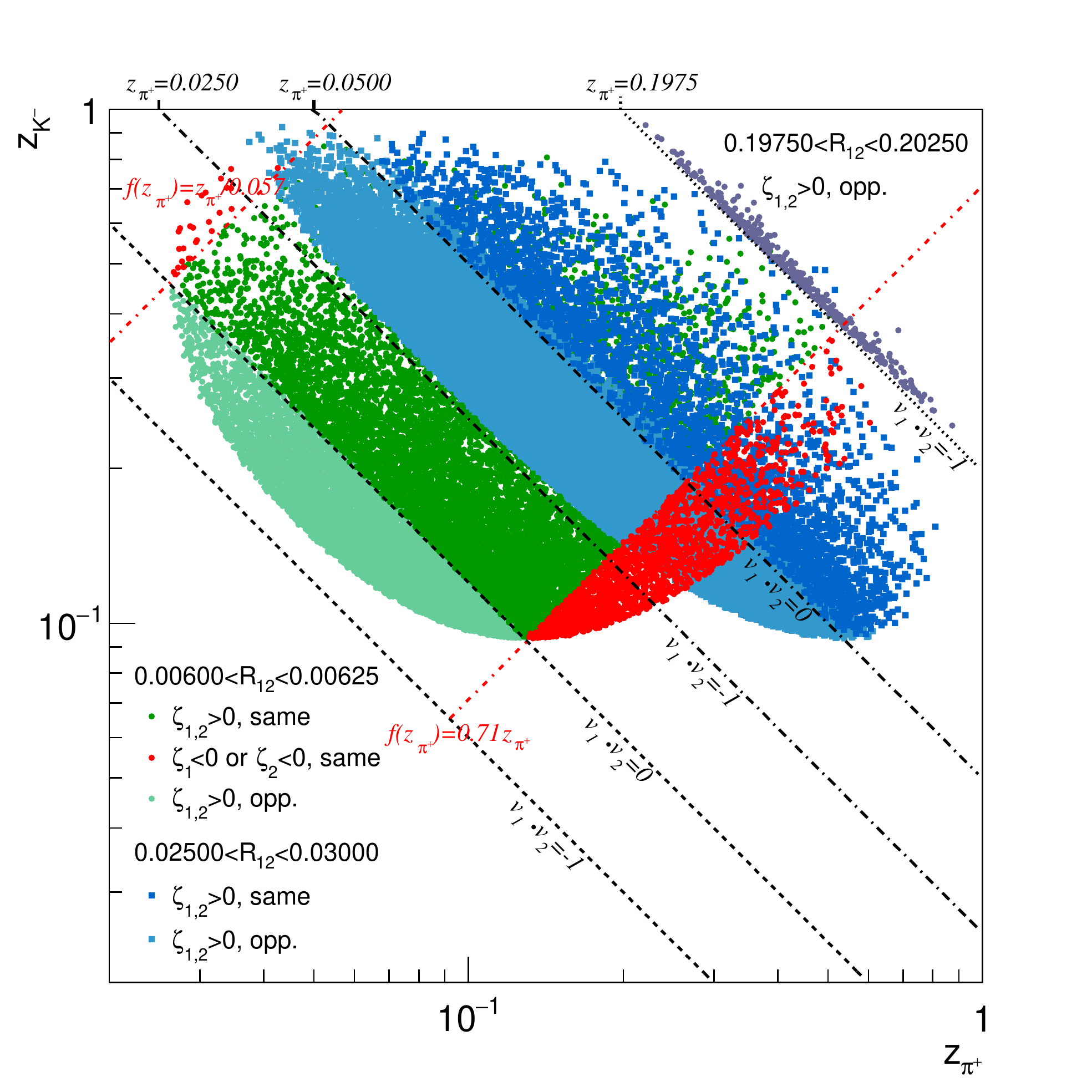,width=1.\textwidth}
\end{minipage}
\caption{As for figure~\ref{fig:scatter}, but for an oppositely charged kaon-pion pair.}
\label{fig:scatter_kaon}
\end{figure*}

\begin{figure*}\centering
\begin{minipage}{0.32\textwidth}\centering
\epsfig{file=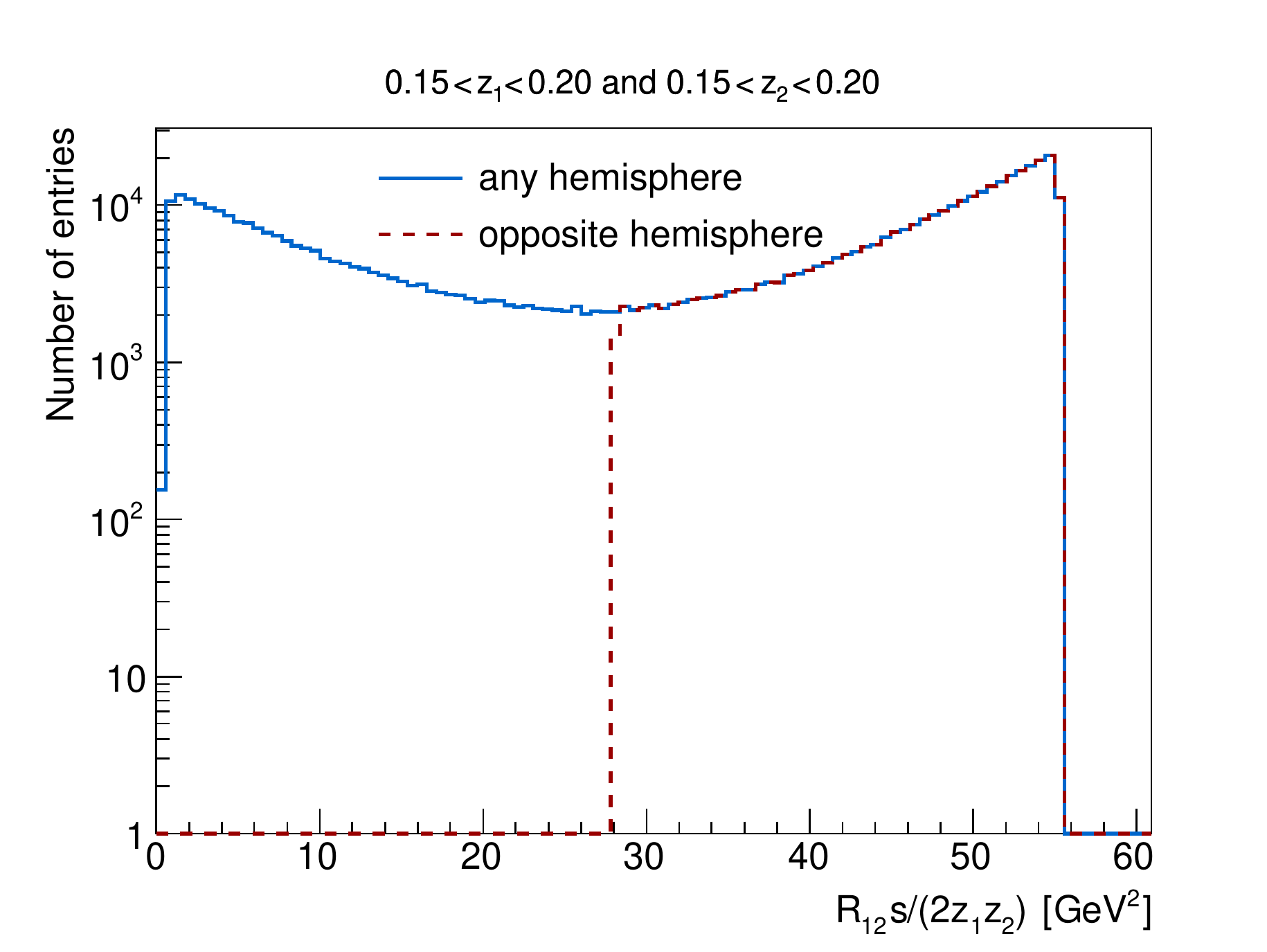,width=1.1\textwidth}
\end{minipage}
\begin{minipage}{0.32\textwidth}\centering
\epsfig{file=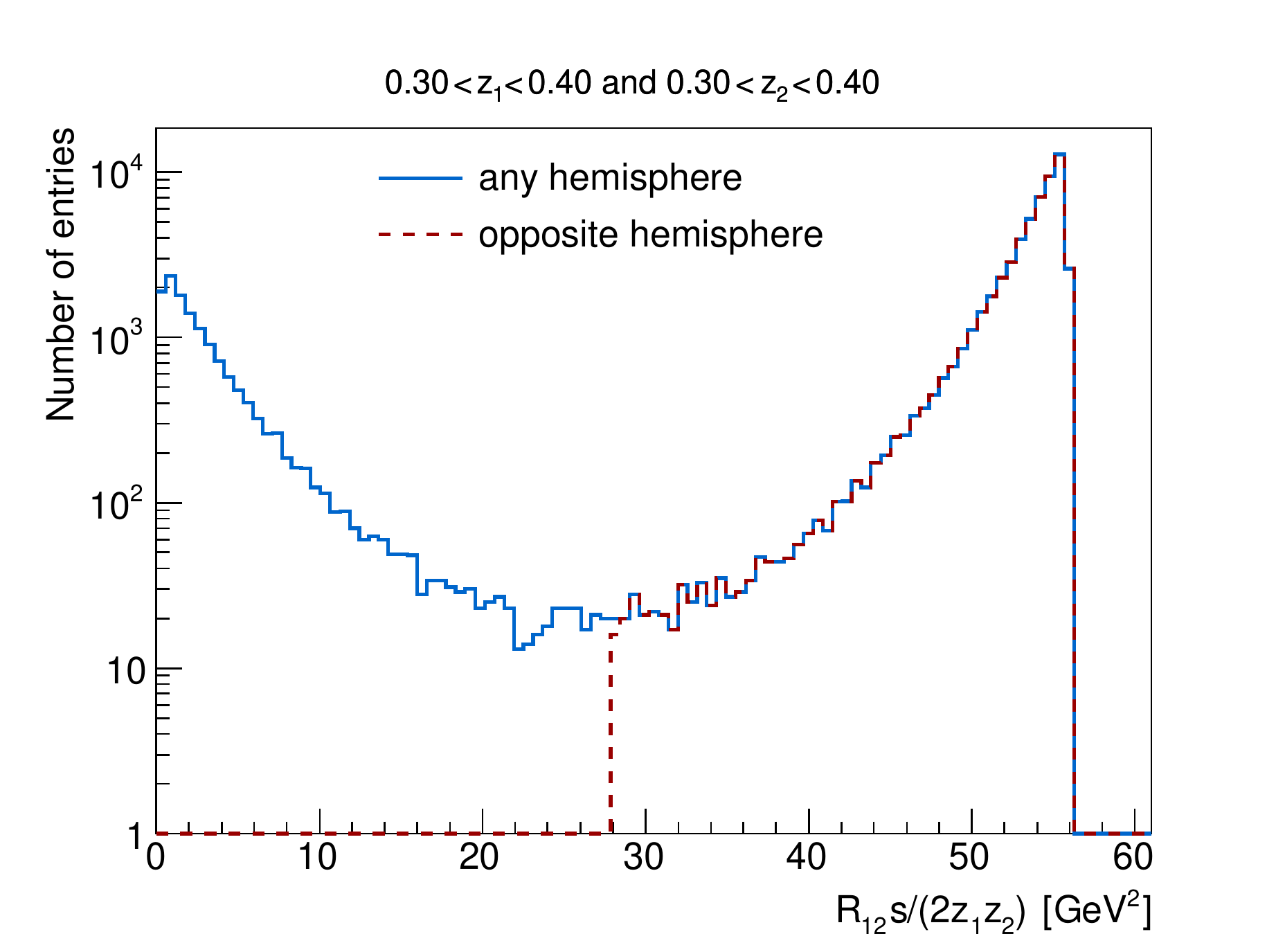,width=1.1\textwidth}
\end{minipage}
\begin{minipage}{0.32\textwidth}\centering
\epsfig{file=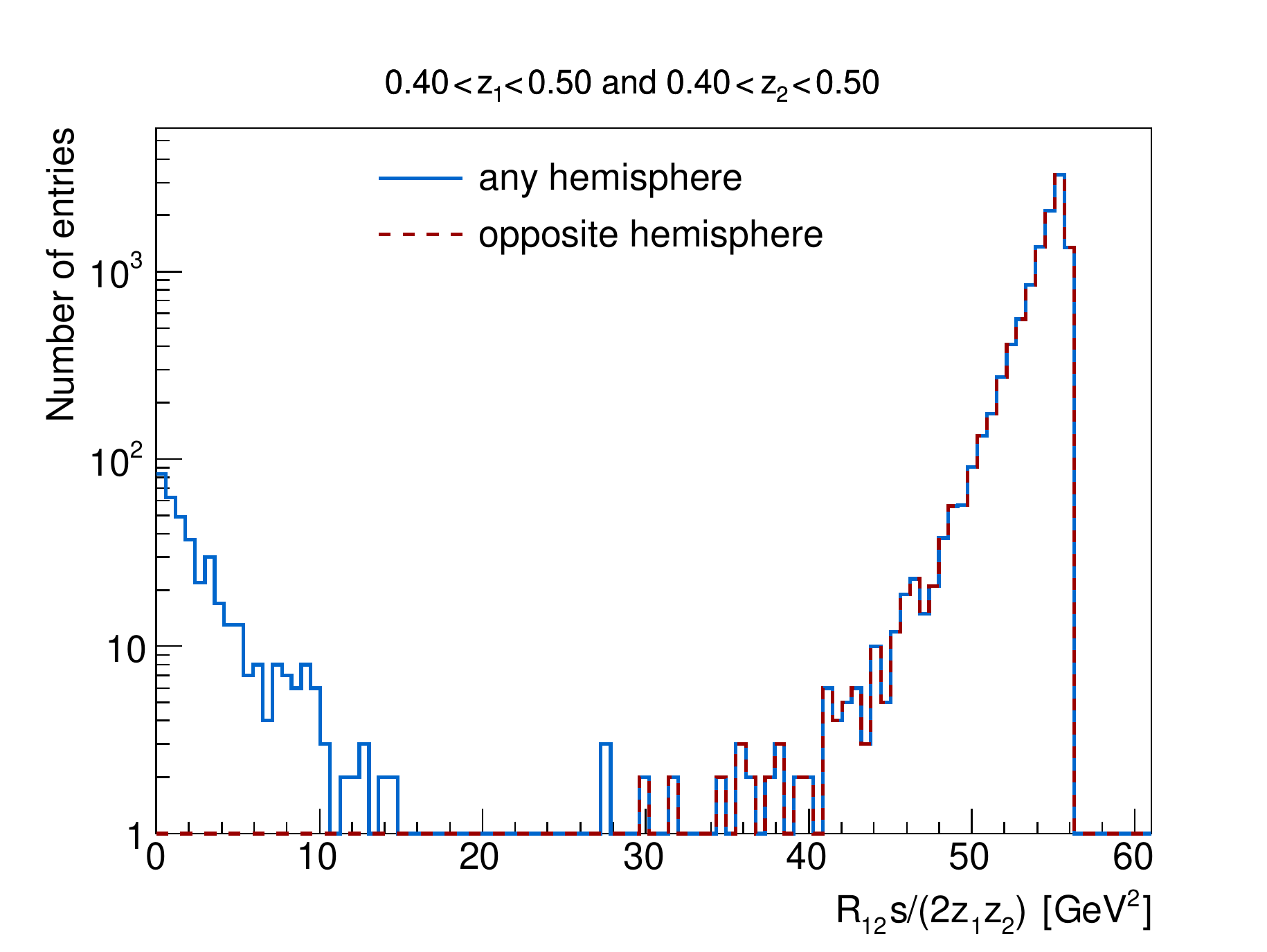,width=1.1\textwidth}
\end{minipage}
\caption{The quantity $R_{12} s/(2 z_1 z_2)$
for $0.15 < z_1 = z_2 < 0.20$ (left), $0.30< z_1 = z_2 < 0.40$ (middle), and $0.40< z_1 =z_2 < 0.50$ (right) for hadrons irrespective of their hemisphere (continuous, blue line), and for hadrons in opposite hemispheres (dashed, red line).} 
\label{fig:pp_zz}
\end{figure*}

For low-energetic hadrons there is an ambiguity in the determination of hemispheres, while  
for hadrons with fractional momenta above 0.1 
both hemisphere selection criteria coincide, as can be seen from the comparison of open and closed symbols for $\zeta_{1}$, $\zeta^{col}_1$, 
and $z_{1}$ in figure~\ref{fig:z_rel_abs_hemi}.
Most commonly, experimental measurements are restricted to fractional hadron momenta above 0.1 because of instrumental limitations, 
while phenomenological extractions of fragmentation functions from experimental data avoid low-$z$ regions because 
of singularities in the QCD evolution equations~\cite{deFlorian:2007aj}.
Figure~\ref{fig:z_rel_abs_hemi} also makes clear that at small values of fractional momentum,  
the three variables $\zeta$, $\zeta^{col}$, and $z$ are relatively close to each other,  
with a good agreement between $\zeta$ and $\zeta^{col}$, while at larger values, they differ more. 

Alternatively, the difference between $\zeta$ and $z$ is also visible in figure~\ref{fig:ktz}.
In this figure, $\sqrt{-q_\st^2}$ is presented (dashed, blue line) together with $\vert P_{2t}\vert /\zeta_{2}$ (continuous, red line) and 
$\vert P_{2t} \vert /z_{2}$ (dash-dotted, green line). Here, $P_{2t}$ refers to the component of hadron 2 orthogonal to $P_1$ of hadron 1, all
measured in the CM frame and with hadron 1 and hadron 2 required to lie in opposite hemispheres.
The results are shown for $\zeta_1=\zeta_2$ comprised between 0.15 and 0.20 (left) and between 0.40 and 0.50 (right). While $\sqrt{-q_\st^2}$
 and $\vert P_{2t}\vert/\zeta_{2}$ (as expected from Eq.~\ref{eq:qTdef}) agree, up to mass corrections, over the full range, both for low and higher regions of
hadron fractional momentum, $\vert P_{2t}\vert/z_{2}$ only agrees with them at low values of $\vert P_{2t}\vert/\zeta_{2}$, 
strongly differing at higher values. This also illustrates the approximate relation in Eq.~\ref{eq:z_zprim_approx}, 
which shows that the difference between $\zeta$ and $z$ increases with increasing $\left|q_\st\right|$. In addition we note that the 
variable $\vert P_{2t}\vert/z_{2}$ is restricted to an upper limit of $\approx \sqrt{s}/2$.

In figure~\ref{fig:scatter}, the distributions of $\zeta^{col}_2$ versus $\zeta^{col}_1$ (left) and 
$z_2$ versus $z_1$ (right) are presented for three ranges in $R_{12}$: $0.00600 < R_{12} < 0.00625$, $0.02500 < R_{12} < 0.03000$, 
and $0.19750 < R_{12} < 0.20250$. 
The respective hemisphere orientation of the hadrons and the sign of $\zeta_i$ are identified by different colours. For the two highest intervals in $R_{12}$, the lower limit of the interval is also indicated in the figures (vertical and horizontal lines). 
The value of $R_{12}$ determines the minimal values that $z_i$ and $\zeta^{col}_i$ can take, as follows from Eqs.~\ref{eq:pp_zz} and~\ref{eq:zzcol_rel}. 
The diagonal lines correspond to $\bm v_1{\cdot}\bm v_2=-1$ or $\bm v_1{\cdot}\bm v_2=0$, as indicated. The latter delimits the configurations where the hadron pairs have the same or the opposite orientation for the respective values of $R_{12}$.
When $\bm v_1{\cdot}\bm v_2=-1$, $\zeta^{col}_1\zeta^{col}_2$ attains its maximum, while 
$z_1 z_2$ reaches its minimum. These products $\zeta^{col}_1\zeta^{col}_2$ and $z_1 z_2$ are then equal to $R_{12}$.

For low enough values of $R_{12}$ and for hadrons in the same hemisphere, $\zeta_i$ can be negative. These events are indicated in red. 
The boundary line for points with $\zeta_i<0$ follows from Eq.~\ref{eq:def_zeta1} or \ref{eq:zeta-relations}. The numerator of Eq.~\ref{eq:zeta-relations} 
 for $\zeta_1$ is always positive, but the denominator turns negative if $\zeta^{col}_2/\zeta^{col}_1 < 2M_2^2/R_{12}s$. This effect is more clearly visible in figure~\ref{fig:scatter_kaon}, where the distribution for a kaon-pion pair is presented and then particularly the region of $\zeta_K < 0$ for smallest $R_{12} \approx 0.006$, in which case the relation is $\zeta^{col}_K/\zeta^{col}_\pi = 2M_K^2/R_{12}s \approx 0.7$. It is asymmetric as for the same $R_{12}$ we have for pions $\zeta^{col}_\pi/\zeta^{col}_K = 2M_\pi^2/R_{12}s \approx 0.06$. Note also that the larger kaon mass creates a distortion of the distributions. The effect is more prominent for the lowest value of $R_{12}$. This region of $R_{12}$ is actually lower than the `safe region' $R_{12} \sim 1$ GeV$^2$/$s$. The analogous situation occurs for the $z_i$ distribution.

For the here presented highest region in $R_{12}$, there are only hadrons originating from opposite hemispheres. 
One has from Eq.~\ref{eq:pp_zz} that for the same-hemisphere configuration, $z_1z_2$ is at its minimum for $\bm v_1{\cdot}\bm v_2=0$. In this limit, and with $R_{12}=0.2$, one obtains  $z_1 z_2=0.4$. For this value of $z_1 z_2$, $z_1+z_2>1$. Hence hadron pairs cannot originate from the same hemisphere, as indeed is observed from figures~\ref{fig:scatter} and~\ref{fig:scatter_kaon}. For values of $R_{12}>0.125$, there are no hadron pairs in the same hemisphere, as follows from Eq.~\ref{eq:pp_zz}.

Equations~\ref{eq:pp_zz} and \ref{eq:pp_zz_cut} are illustrated in figure~\ref{fig:pp_zz}, where 
$R_{12} s/(2 z_1 z_2)$ is shown for increasing ranges of $z_1=z_2$ for hadron pairs irrespective of their hemisphere (continuous, blue line) and for hadron pairs in opposite hemispheres (dashed, red line). 
At $z_1 z_2=2 R_{12}$, i.e., $R_{12} s/(2 z_1 z_2)=s/4=27.98$~GeV$^2$, there is a sharp separation between the regions corresponding to 
the same-hemisphere and opposite-hemisphere configuration. 
At low values of fractional momentum, the transition between these regions is continuous, while for higher fractional momenta, both regions 
are progressively better separated from each other. With increasing fractional momenta, there are also gradually less pairs of hadrons belonging to the 
same hemisphere, while at the highest values of $z_i$, their contribution completely disappears.

\begin{figure*}\centering
\begin{minipage}{0.48\textwidth}\centering
\epsfig{file=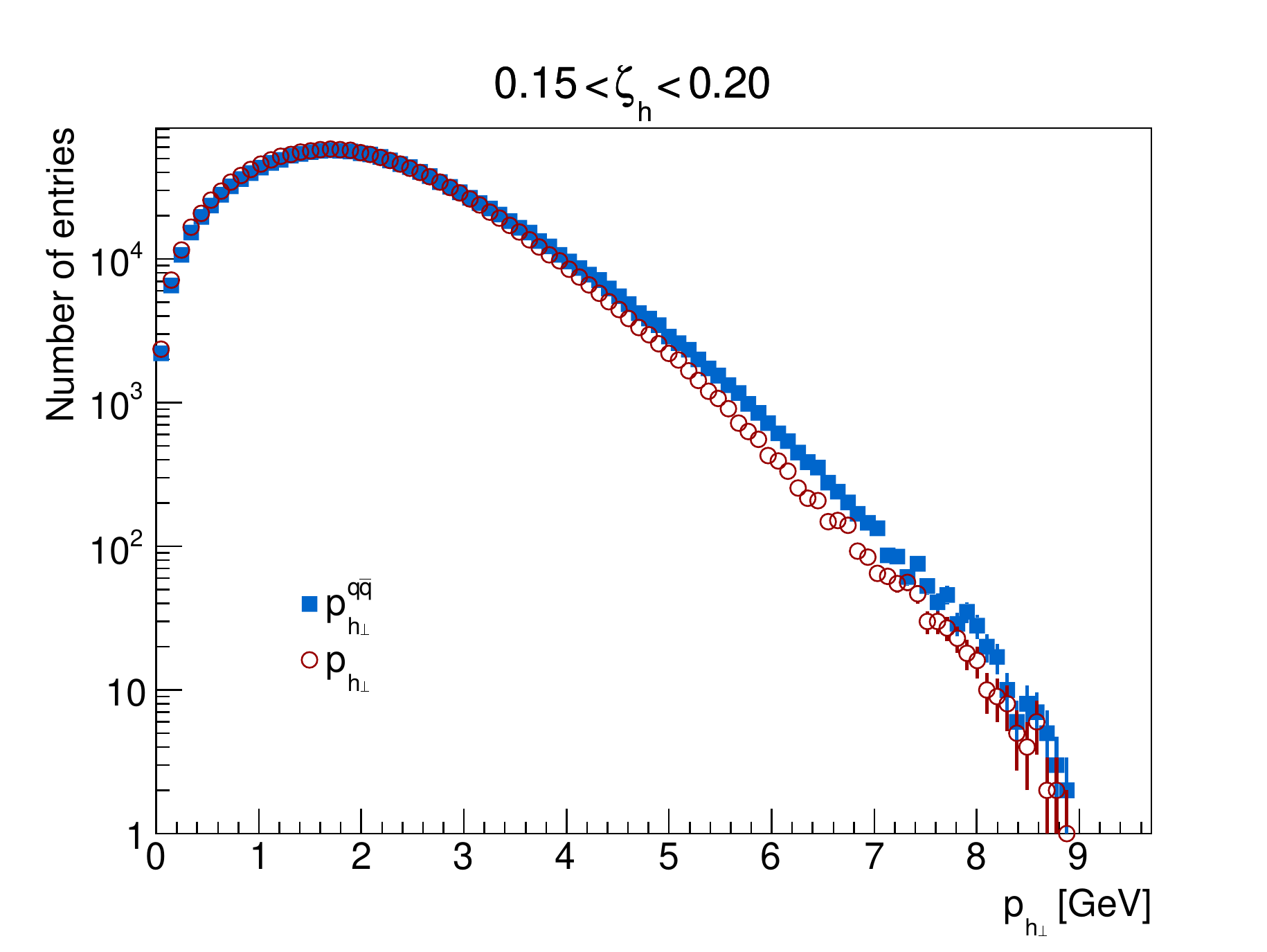,width=1.\textwidth}
\end{minipage}
\begin{minipage}{0.48\textwidth}\centering
\epsfig{file=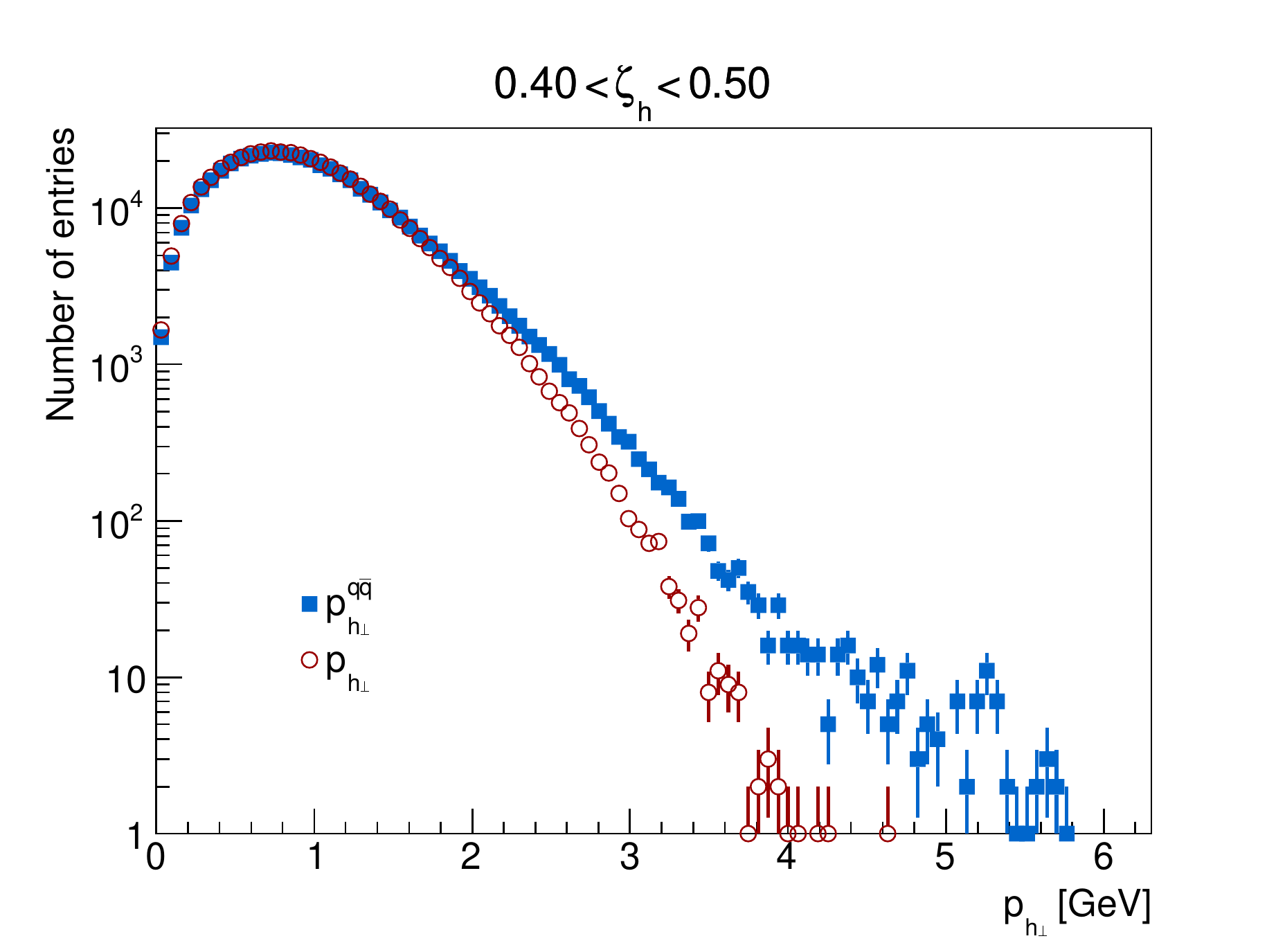,width=1.\textwidth}
\end{minipage}
\caption{The scaled transverse momentum $p_{h\perp}$ extracted from $q_\st$ (red, open circles), using the thrust axis as reference, and the transverse momentum of hadron $h$ with respect to the $q\bar{q}$ axis, scaled with $\zeta_h$ (blue, closed squares) for $0.15<\zeta_{h}<0.20$ (left) and $0.40<\zeta_{h}<0.50$ (right).} 
\label{fig:k2T}
\end{figure*}

\begin{figure*}\centering
\begin{minipage}{0.48\textwidth}\centering
\epsfig{file=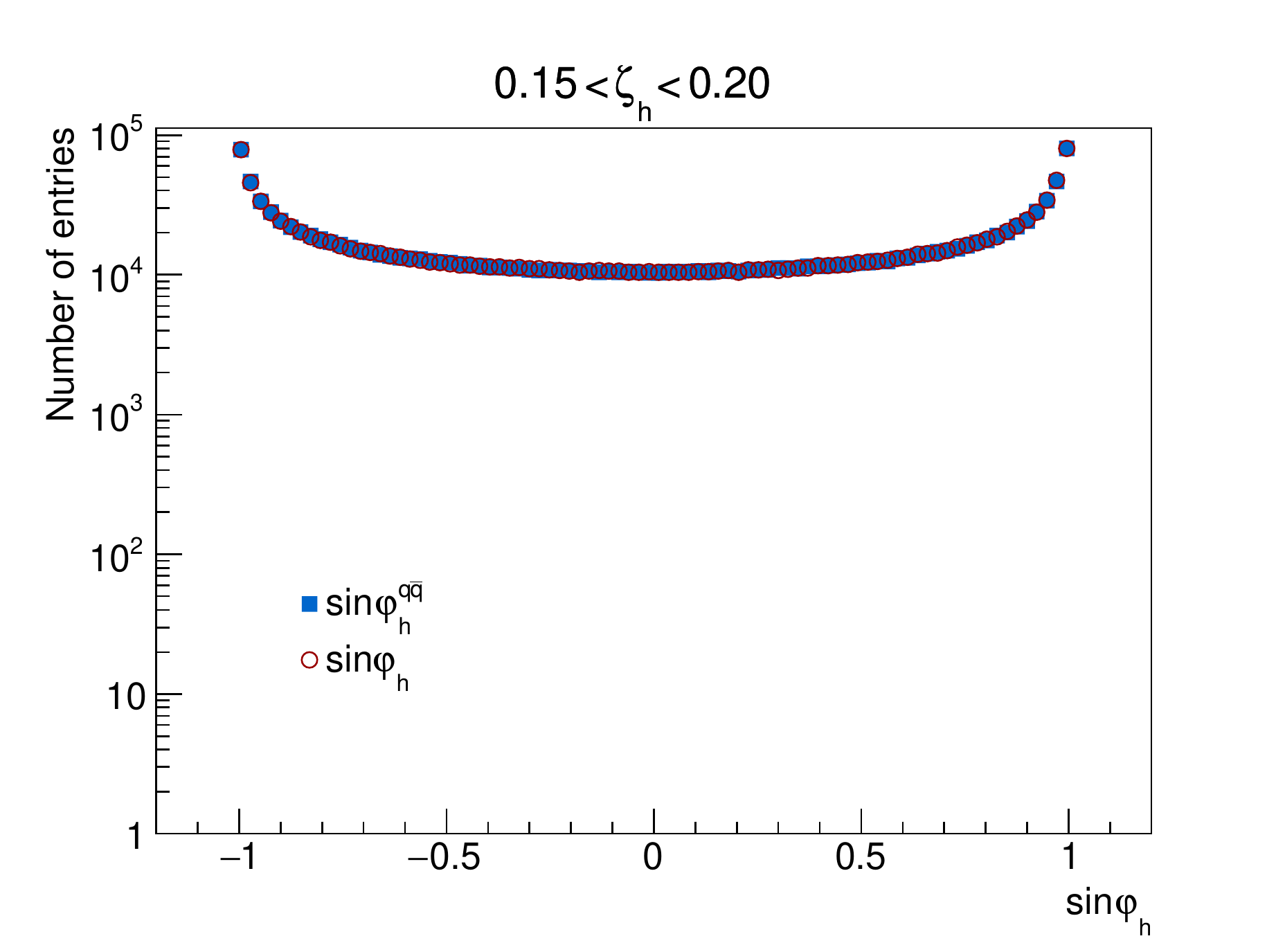,width=1.\textwidth}
\end{minipage}
\begin{minipage}{0.48\textwidth}\centering
\epsfig{file=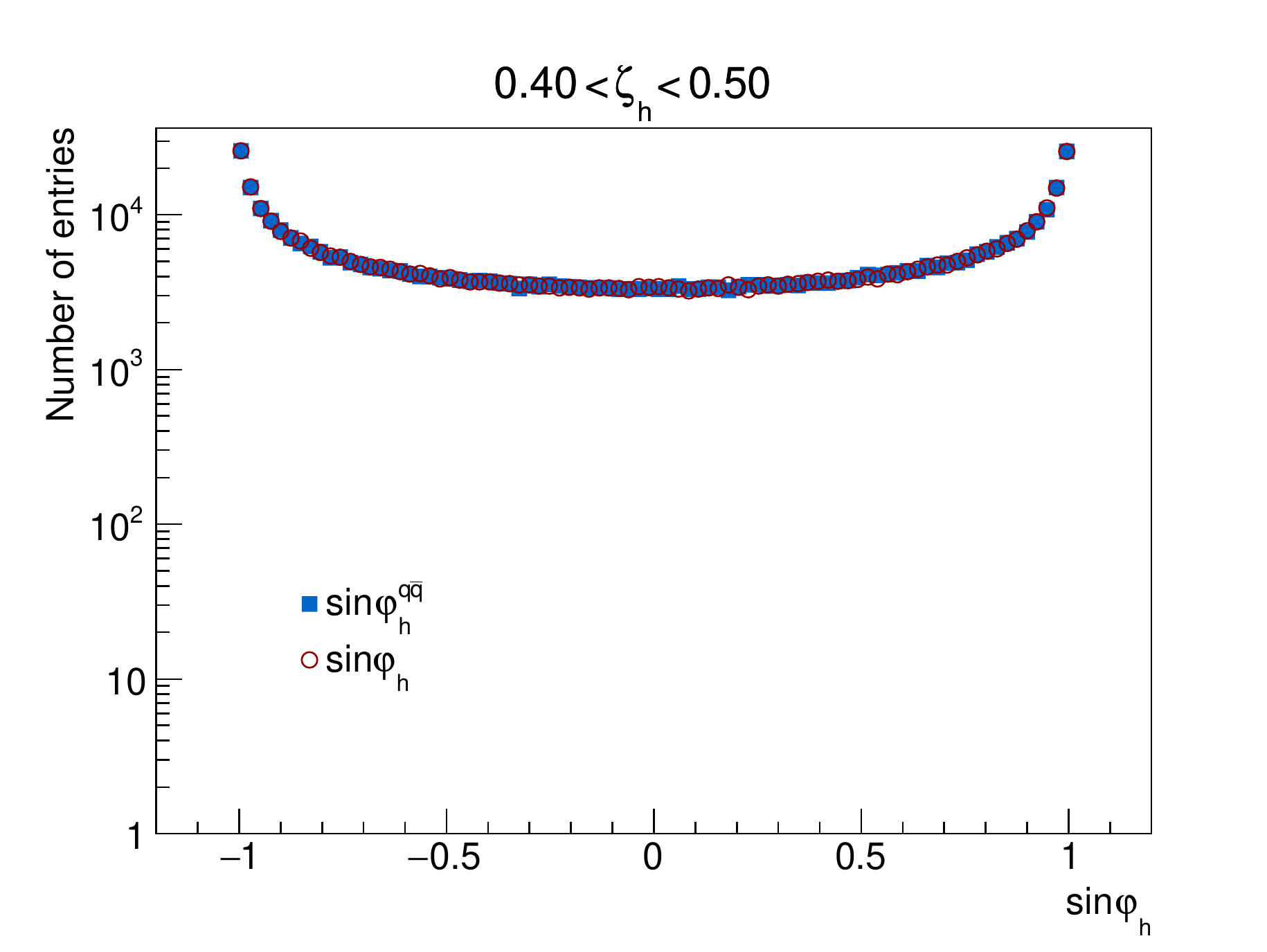,width=1.\textwidth}
\end{minipage}
\caption{The quantity $\sin\varphi_{h}$ extracted from $D_\st$ (red, open circles), using the thrust axis as reference, and $\sin\varphi_{h}^{q\bar{q}}$ of hadron $h$ with respect to the plane formed by the lepton and $q\bar{q}$ axis (blue, closed squares) for $0.15<\zeta_{h}<0.20$ (left) and $0.40<\zeta_{h}<0.50$ (right).} 
\label{fig:sinphi}
\end{figure*}

In figure~\ref{fig:k2T}, $\vert p_{h\perp} \vert$ obtained from Eq.~\ref{qT2-2} (open, red circles), 
using the thrust axis as reference axis, is compared to the hadron transverse-momentum component with respect to the $q\bar{q}$ axis 
scaled with $\zeta_h$, $p_{h\perp}^{q\bar{q}}$, (filled, blue squares), for $0.15<\zeta_h<0.20$ (left) and $0.40<\zeta<0.50$ (right). 
A lower limit on the value of the thrust variable of 0.85 is required in order to select events that have a two-jet configuration.
This cut affects the distribution of the transverse momentum, since it affects to which degree tracks are collimated rather 
than isotropically distributed. 
At small values of $\vert p_{h\perp} \vert$, 
there is good agreement between the generated distribution and that obtained from Eq.~\ref{qT2-2}. 
At larger values disagreement is observed, which can be explained by the contribution of events where more than two jets are created. 

In figure~\ref{fig:sinphi}, $\sin\varphi_{h}$ extracted from Eq.~\ref{DT-2} (open, red circles), 
using the thrust axis as reference, with as explained for Eqs.~\ref{qT2-2} and~\ref{DT-2}, 
$\zeta$ of the thrust axis set to 1 and the magnitude of the thrust axis normalised to $\sqrt{s}/2$, is compared to 
$\sin\varphi_{h}^{q\bar{q}}$, where the angle $\varphi_{h}^{q\bar{q}}$ is the 
angle between the plane containing the $q\bar{q}$ axis and the $e^+e^-$ axis and the plane containing the $q\bar{q}$ axis and hadron $h$. 
The $\zeta_h$ ranges covered are the same as those in figure~\ref{fig:k2T} and also here a minimal thrust value of 0.85 is required. 
The distribution is insensitive to the latter requirement.
The $\sin\varphi_h$ and  $\sin\varphi_{h}^{q\bar{q}}$ distributions show a very good agreement, irrespective of the value of $\zeta_{h}$.
Extracting the determinant defined in Eq.~\ref{eq:DT_def}, using as reference the thrust axis, offers thus the possibility to learn about the hadron azimuthal distribution.

\begin{figure*}\centering
\begin{minipage}{0.48\textwidth}\centering
\epsfig{file=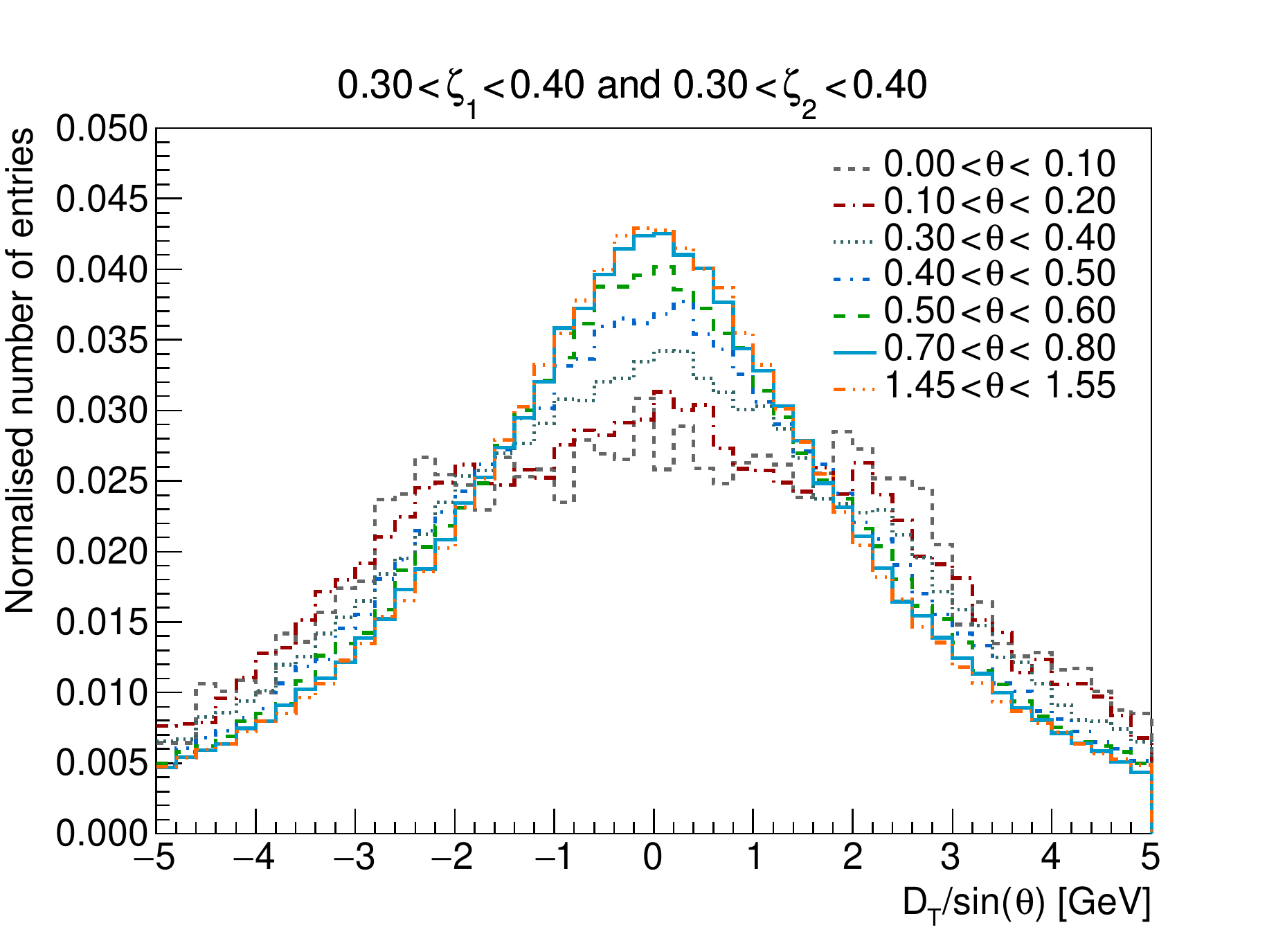,width=1.\textwidth}
\end{minipage}
\begin{minipage}{0.48\textwidth}\centering
\epsfig{file=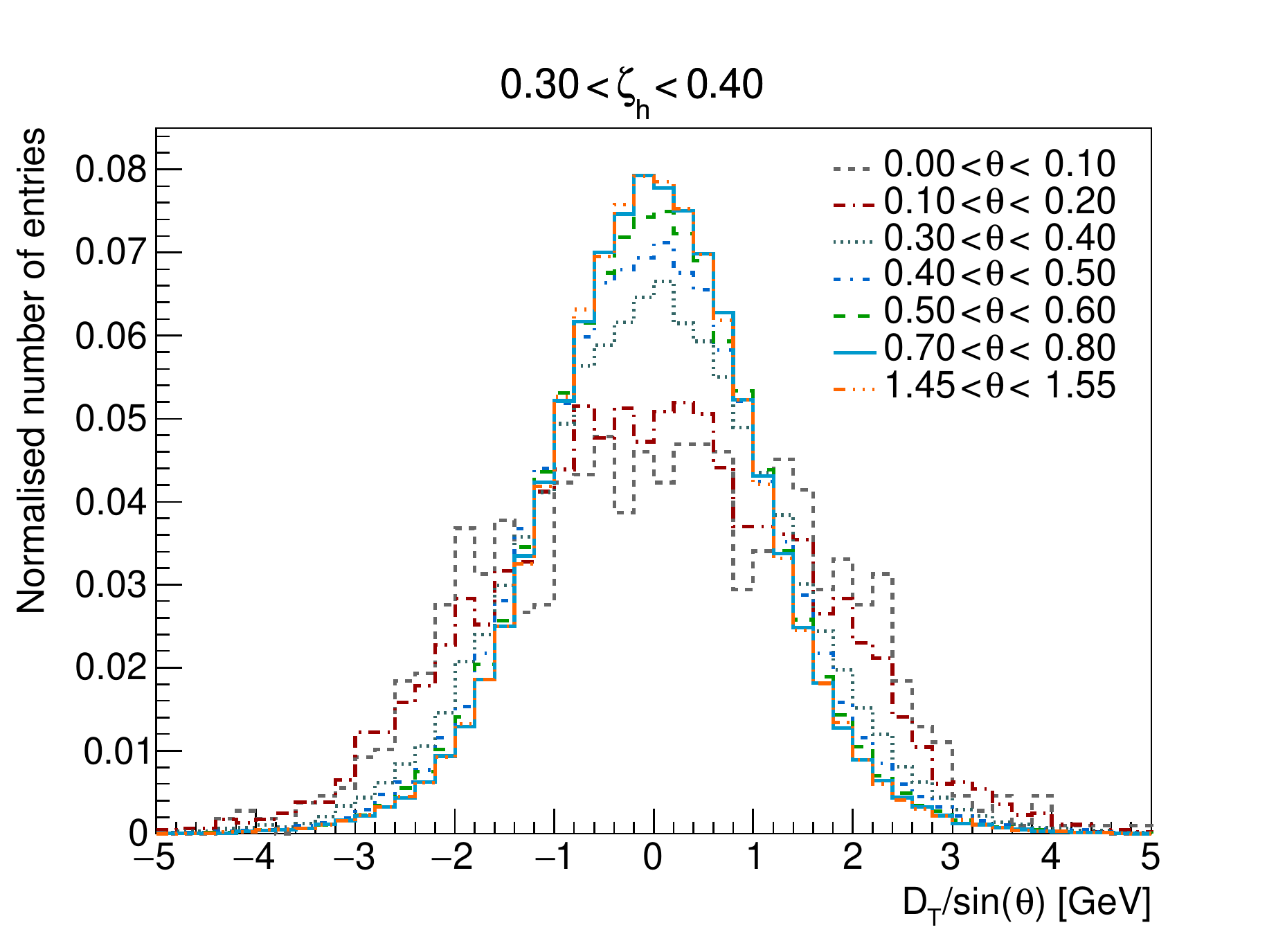,width=1.\textwidth}
\end{minipage}
\caption{The distribution of $D_\st/\sin\theta$ for the event configuration considering pairs of hadrons from opposite hemispheres (left) and the hadron--thrust-axis  combination (right),  for $0.3<\zeta_{1,2(h)}<0.4$ and for different ranges of $\sin\theta$, as indicated.}
\label{fig:det_sin}
\end{figure*}

To illustrate the $D_\st$ distribution by itself, we begin by recalling that it is not uniformly distributed in $\theta$. Firstly, the $\theta$-dependence is determined by the hard process, where we expect the cross section of $e^+e^-\rightarrow q\bar q$ to have a $(1 + \cos^2\theta)$ behavior. This dependence, ideally, is decoupled from the non-collinearity due to the fragmentation process. Secondly, there is the $\sin\theta$ and $\cos\theta$ dependence that comes back in the orientation of the hadron production planes as defined in figure~\ref{eeqq12} and appears in Eq.~\ref{DT-1}. In particular close to $\theta = 0$ and $\theta = \pi$, there is a contribution of the second term in Eq.~\ref{DT-1}, even if its magnitude is suppressed by a factor $\vert p_\perp\vert/\sqrt{s} \lesssim 0.1$ as compared to the first term. Note, however, that for $\theta = 0$ and $\theta = \pi$, the azimuthal angle is not defined. 
In figure~\ref{fig:det_sin},  the distribution of $D_\st/\sin\theta$ is shown for two hadrons from opposite hemispheres (left) and for the configuration of a hadron and 
the thrust axis (right).
Here,  
\bea 
\sin\theta &=& \frac{2\sqrt{tu}}{s} = \frac{4\sqrt{(l_1{\cdot}k_1)(l_2{\cdot}k_1)}}{s}  
\eea
is obtained from the directions in the CM frame of the lepton axis and a (fast) hadron or the thrust axis:
\bea
\sin\theta \approx \frac{\arrowvert l_1\times P_{1} \arrowvert}{\arrowvert l_1 \arrowvert \arrowvert P_{1}\arrowvert},
\label{eq:sin_ap}
\eea
where $P_1$ is to be replaced by $P_{\rm thrust}$ when considering the thrust axis.
Other possible approximations for $\sin\theta$, in order of decreasing accuracy, involve one or two of the hadron momenta,
\bea
\sin\theta &\approx& \frac{4\sqrt{(l_1{\cdot}P_{1})(l_2{\cdot}P_{1})}}{\zeta_{1}\,s} \\
& \approx& \frac{4\sqrt{(l_1{\cdot}P_{1})(l_1{\cdot}P_{2})}}{s\sqrt{\zeta_{1}\zeta_{2}}},
\eea
where the hadron $P_1$ in the first equation or one of the hadrons in the second equation can be replaced by $P_{\rm thrust}$ (and also $\zeta_{\rm thrust}$), 
which gives a better accuracy. In particular, replacing the hadron in the first approximation by the thrust axis coincides with Eq.~\ref{eq:sin_ap} for the thrust axis.
The $D_\st/\sin\theta$ distribution shows a dependence on $\theta$ for low and mid values of $\theta$, whereas for higher values, 
where the $\cos\theta$ contribution in $D_\st$ is strongly suppressed, no $\theta$ dependence is observed.
The distribution from two opposite hemisphere hadrons nicely shows a transverse momentum distribution with a width of the order of $1.5$ GeV. Using the thrust axis and one hadron the width is about $1.0$ GeV, indeed something like a factor $1/\sqrt{2}$ less as expected for a uniform transverse momentum distribution.

\section{Conclusions}

Two measures for non-collinearity have been introduced and discussed. 
One is the invariant length of the non-collinearity vector $q_\st$, in a back-to-back jet situation directly related to the transverse momenta of partons in the hadrons that are produced in the hadronisation process.
The other is a determinant constructed from the two lepton momenta and the hadron and thrust momenta, or from the two lepton momenta and two opposite-hemisphere--hadron momenta, which in the back-to-back jet situation contains information on the transverse momentum orthogonal to the lepton-parton plane. The latter thus also contains information on the azimuthal orientation of the back-to-back partons through the orientation of the momenta of the produced hadrons with respect to the lepton plane. 

Our results show that the 2PI variables ($\zeta$) are perfectly suitable to correct for various shortcomings of the variables $\zeta^{col}$ or the CM energy fractions $z$. The 2PI variables $\zeta$ by construction account for hadronic mass corrections and the deviation from non-collinearity. They are not constrained between 0 and 1 as the other variables, but in the back-to-back jet situation the $\zeta$'s also tend to approximate light-cone momentum fractions (just as $\zeta^{col}$), which are suitable variables when it comes to QCD factorisation. 

In how far the given measures for non-collinearity, $q_\st$ and $D_\st$, measure the transverse momenta for an individual fragmenting parton depends on the appropriate definition of these quantities in a high energy scattering process. We considered in this note their importance as approximants of quantities that have a natural interpretation in the back-to-back two-jet situation.

All quantities and corresponding distributions have been exemplified with studies from a Monte-Carlo simulation. Alongside, three different momentum fractions have been presented and compared, and their relation to quantities characterising hemisphere orientation and the transverse momentum acquired in the hadronisation process has been exposed.

\section*{Acknowledgements}

We acknowledge useful discussions with several colleagues, among them Dani\"el Boer, Alessandro Bacchetta and Marco Radici. 
We are grateful to Ralf Seidl for the Monte-Carlo simulation. 
PJM acknowledges the support of the FP7 EU "Ideas" programme QWORK (Contract 320389).
CVH acknowledges the support from the Basque Government (Grant No. IT956-16) and the
Ministry of Economy and Competitiveness (MINECO) (Juan de la Cierva), Spain.

\bibliographystyle{apsrev}

\end{document}